\documentclass[aps,a4paper,twocolumn]{revtex4-2}
\usepackage{amsmath}
\usepackage{graphicx}
\usepackage{subfigure}
\usepackage[usenames,dvipsnames]{color}
\definecolor{darkblue}{RGB}{0,0,196}
\usepackage[colorlinks=true,linkcolor=darkblue,citecolor=darkblue,urlcolor=darkblue]{hyperref}
\usepackage{setspace}
\usepackage{hyperref}
\usepackage{xcolor}
\hypersetup{
  colorlinks   = true, 
  urlcolor     = red, 
  linkcolor    = blue, 
  citecolor   = blue 
}
\usepackage{graphicx}
\usepackage{amsmath,bbm}
\usepackage{amssymb,bm}
\usepackage{yfonts}
\usepackage{comment}

\begin{document}

\title{Anisotropic flow fluctuation as a possible signature of clustered nuclear geometry\\ in O--O collisions at the Large Hadron Collider 
}


\author{Suraj Prasad}\email[]{Suraj.Prasad@cern.ch}
\author{Neelkamal Mallick}\email[]{Neelkamal.Mallick@cern.ch}
\author{Raghunath Sahoo}\email[Corresponding Author: ]{Raghunath.Sahoo@cern.ch}
\affiliation{Department of Physics, Indian Institute of Technology Indore, Simrol, Indore 453552, India}

\author{Gergely G\'abor Barnaf\"oldi}\email[]{barnafoldi.gergely@wigner.hun-ren.hu}
\affiliation{Wigner Research Center for Physics, 29-33 Konkoly-Thege Mikl\'os Str., H-1121 Budapest, Hungary}

\begin{abstract}
Nuclei having $4n$ number of nucleons are theorized to possess clusters of $\alpha$ particles ($^4$He nucleus). The Oxygen nucleus ($^{16}$O) is a doubly magic nucleus, where the presence of an $\alpha$-clustered nuclear structure grants additional nuclear stability. In this study, we exploit the anisotropic flow coefficients to discern the effects of an $\alpha$-clustered nuclear geometry with respect to a Woods-Saxon nuclear distribution in O--O collisions at $\sqrt{s_{\rm NN}}=7$ TeV using a hybrid of IP-Glasma + MUSIC + iSS + UrQMD models. In addition, we use the multi-particle cumulants method to measure anisotropic flow coefficients, such as elliptic flow ($v_{2}$) and triangular flow ($v_{3}$), as a function of multiplicity class. Anisotropic flow fluctuations, which are expected to be larger in small collision systems, are also studied for the first time in O--O collisions. It is found that an $\alpha$-clustered nuclear distribution gives rise to an enhanced value of $v_{2}$ and $v_3$ for the low-multiplicity events. Consequently, a rise in $v_3/v_2$ is also observed for the 0--10\% multiplicity class. Further, for $\alpha$-clustered O--O collisions, fluctuations of $v_{2}$ are larger for the highest multiplicity events, which decrease as the final-state multiplicity decreases. In contrast, for a Woods-Saxon $^{16}$O nucleus, $v_{2}$ fluctuations show an opposite behavior with decreasing multiplicity. When confronted with experimental data, this study may reveal the importance of the nuclear density profile on the discussed observables and provide physics validation for the hybrid model discussed in this work.
\end{abstract}
\date{\today}
\maketitle 

\section{Introduction}
\label{intro}

The primary goal of relativistic heavy-ion collisions at the RHIC and the LHC is to create a deconfined thermalized medium of partons, also known as Quark Gluon Plasma (QGP), which is believed to have existed a few microseconds after the Big Bang. However, due to the transient nature of QGP, it can not be observed directly in experiments. Instead, several indirect signatures from experimentally measurable observables can signify the presence of such a deconfined strongly interacting matter in heavy-ion collisions. One such observable is azimuthal anisotropy, which characterizes the collective behavior of the medium formed in heavy-ion collisions. It is quantified in terms of the Fourier expansion coefficients of azimuthal distribution of final state charged hadrons, as follows~\cite{Voloshin:1994mz}:
\begin{equation}
\frac{dN}{d\phi}\propto \Bigg(1 + 2\sum_{n=1}^{\infty}v_{n}\cos\left[n(\phi-\psi_{n})\right]\Bigg) \ . 
    \label{eq:fourierexpansion}
\end{equation}
Here, $\phi$ is the azimuthal angle, $\psi_{n}$ stands for the $n$\textsuperscript{th} order symmetry plane angle and $v_{n}$ being the $n$\textsuperscript{th} order flow coefficients. Here, $v_{2}$ quantifies the elliptic flow, $v_{3}$ gives the estimate for triangular flow, and so on. Both $v_{2}$ and $v_{3}$ depend upon the transport properties of the QCD medium formed during heavy-ion collisions. In addition, studies based on hydrodynamic calculations show that the flow coefficients are affected strongly by a change in the ratio of shear viscosity to entropy density ($\eta/s$)~\cite{Schenke:2011bn, Chaudhuri:2011pa}. Moreover, the sensitivity of flow coefficients to $\eta/s$ is large for higher-order harmonics, i.e., $v_{3}$ is more sensitive to $\eta/s$ as compared to $v_2$~\cite{Chaudhuri:2011pa}. In several studies, $v_{2}$ and $v_{3}$ are found to be associated with the initial eccentricities of the participating nucleons, where the values of anisotropic flow coefficients can change when the nuclear distribution is modified~\cite{Behera:2023nwj, Giacalone:2021udy, Haque:2019vgi}. In addition, in Xe--Xe collisions, an observation of a larger value of $v_2$ in the most central collisions, as compared to that observed in Pb--Pb collisions, is anticipated due to a deformed nuclear structure of Xe nucleus~\cite{ALICE:2021ibz, ALICE:2018lao, ATLAS:2019dct, CMS:2019cyz}. In a similar line, the nuclear deformation of Uranium, being a prolate shape nucleus, is reflected in a higher $v_{2}$, when compared to that observed in the most central collisions of Au nuclei, which is close to a spherical nucleus~\cite{STAR:2015mki}. Similarly, one notices that the values of $v_{2}$ and $v_{3}$ vary with the change in the deformation parameters of Uranium in U--U collisions, and the effect is enhanced for $v_{2}$, as compared to $v_{3}$~\cite{Giacalone:2021udy, Haque:2019vgi, PHENIX:2018lia}.
\begin{figure*}
\begin{center}
\includegraphics[scale=0.5]{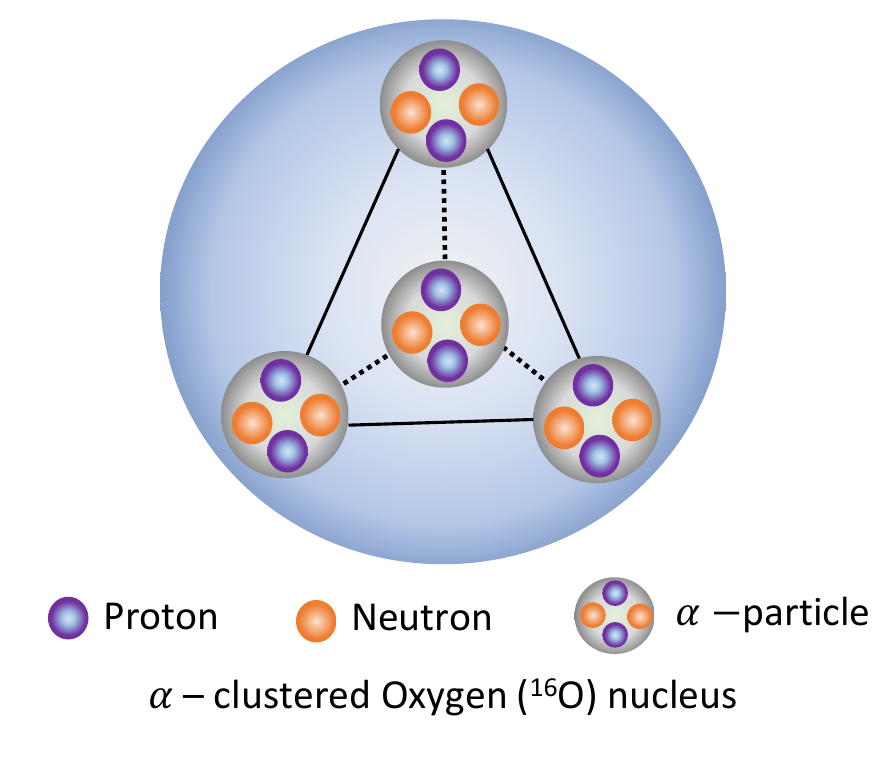}
\includegraphics[scale=0.5]{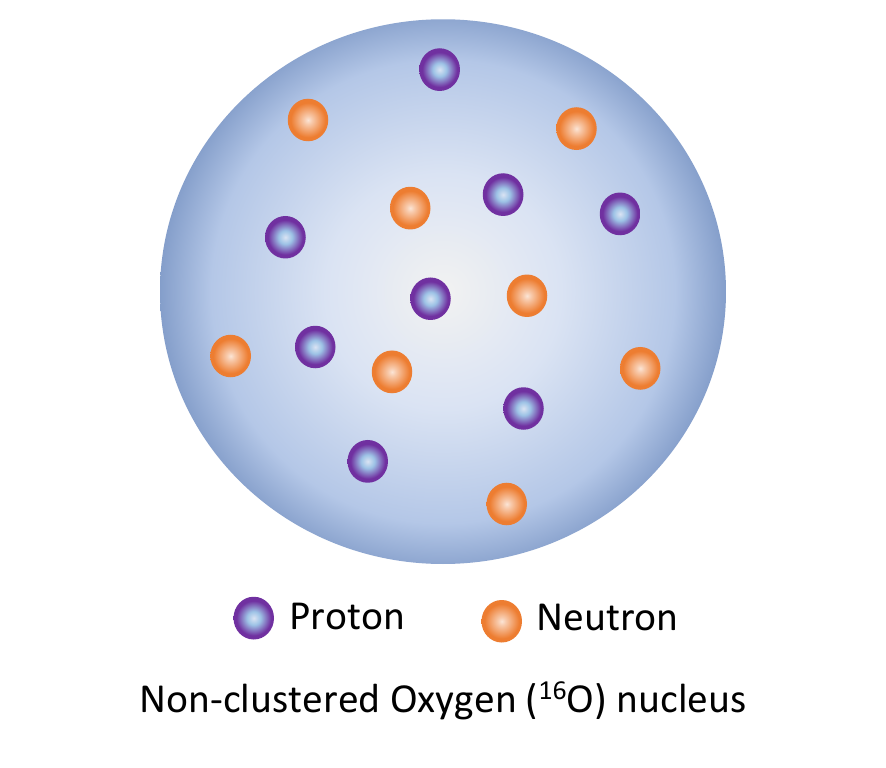}
\caption{Pictorial representation of an $\alpha$-clustered (left) and non-clustered (right) distribution of nucleons inside the $^{16}$O nucleus.}
\label{fig:picalphaWS}
\end{center}
\end{figure*}

Elliptic flow is one of the observables sensitive to the initial state effects in heavy-ion collisions~\cite{Gardim:2011xv}. However, the value of elliptic flow fluctuates from one event to another, which is a reflection of the fluctuation in the initial energy density and event-by-event participant nucleon distribution before the formation of QGP~\cite{ALICE:2022zks}. These fluctuations in the initial density profile of the participating nucleons can generate odd flow harmonics and lead to fluctuations in the harmonic flow coefficients~\cite{ALICE:2022zks}. Further, the fluctuations in the initial energy density can develop symmetry planes of different orders in various kinematic regions~\cite{CMS:2015xmx, ALICE:2017lyf, ATLAS:2017rij}. In other words, the initial density fluctuations can develop transverse momentum ($p_{\rm T}$) and pseudorapidity ($\eta$) dependence of $\psi_{n}$. Therefore, to comprehend the event-by-event initial density fluctuations and their impact on medium evolution, the study of anisotropic flow fluctuations becomes crucial. Anisotropic flow fluctuations are well explored in both experimental and theoretical frontiers of heavy-ion collisions~\cite{PHOBOS:2007vdf, PHOBOS:2010ekr, PHENIX:2018lfu, ALICE:2018yph, ALICE:2012vgf, ALICE:2014wao, ALICE:2016cti}. The measurements by the ATLAS experiment hint at the non-Gaussian behavior of the flow fluctuations in Pb--Pb collisions~\cite{ATLAS:2013xzf}. Consequently, this observation has led to studies constraining their probability distribution functions~\cite{CMS:2017glf}. However, the anisotropic flow fluctuations in small collision systems are not yet well understood. Moreover, one may conceive that for a small number of participants, the contribution to fluctuations can increase. Finally, if the evolution of anisotropic flow fluctuations has more than one contributor, then the results can vary significantly from the expectations.

In addition to the collective behavior of the QGP, several other signatures for the presence of a QGP medium have been observed in heavy-ion collisions. For the measurement of such signatures, the baseline measurements are obtained from proton-proton ($pp$) collisions, where the formation of a QCD medium is not anticipated. However, recent observations of strangeness enhancement~\cite{ALICE:2016fzo}, ridge-like structure~\cite{CMS:2015fgy, ALICE:2013snk}, radial flow effects~\cite{ALICE:2013wgn, ALICE:2016dei, CMS:2016fnw}, etc., in high multiplicity $pp$ collisions have highlighted the significance of studying the small collision systems from different aspects. 
In addition, recent measurements with ALICE confirm the observation of partonic collectivity in $pp$ and p--Pb collisions, where a characteristic grouping (within $\sim 1\sigma$) and splitting (within $ \sim 5 \sigma$) of elliptic flow for mesons and baryons at intermediate $p_{\rm T}$ is observed~\cite{ALICE:2024vzv}. The LHC and RHIC have plans to inject Oxygen ($^{16}$O) nuclei to perform p--O and O--O collisions~\cite{Brewer:2021kiv, Katz:2019qwv}, which can bridge the multiplicity gap between $pp$, p--Pb and Pb--Pb collisions and may provide insight into the possible formation of QGP droplets in small systems. As already mentioned, the presence of collective transverse expansion of the medium (also known as collectivity) in ultrarelativistic collisions of nuclear matter is regarded as one of the key signatures of QGP.  This leads to the appearance of long-range multi-particle azimuthal correlations in the final state. Through many extensive studies, it has been understood that this collective expansion can be modeled via relativistic hydrodynamics with dissipative effects, which is responsible for these long-range correlations observed in heavy-ion collisions. By comparing the hydrodynamics model predictions with experimentally observed azimuthal anisotropy, it confirms that the initial-state effects such as eccentricity, nuclear deformation, and density fluctuations in the colliding species get embedded in the final-state multi-particle correlations through this collective medium evolution and manifest as finite measurable flow coefficients. Based on these observations from heavy-ion collisions, the existence of long-range multi-particle correlations in high multiplicity events in small collision systems can strongly be associated with the formation of an early deconfined and locally thermalized partonic phase in small collision systems such as O--O collisions.

Furthermore, $^{16}$O is a stable nucleus having a double magic number and is theorized to possess clusters of $\alpha$ ($^4$He) particles, where the $\alpha$-particles arrange themselves in the corners of a tetrahedron~\cite{gamow, Wheeler:1937zza, Bijker:2014tka, Wang:2019dpl, He:2014iqa, He:2021uko, Otsuka:2022bcf}. Figure~\ref{fig:picalphaWS} shows a pictorial representation of the nucleons inside an $^{16}$O nucleus for an $\alpha$-clustered structure (left) and non-clustered distribution of nucleons (right). With the RHIC geometry scan program, $^{16}$O collisions would be an interesting study which may unveil the inner geometry of $^{16}$O nuclei~\cite{PHENIX:2018lia, PHENIX:2021ubk, STAR:2022pfn, STAR:2023wmd}. In recent years, there have been several studies performed to understand the O--O collisions both at RHIC and LHC energies~\cite{Bozek:2014cva, Broniowski:2013dia, Li:2020vrg, Rybczynski:2019adt, Sievert:2019zjr, Huang:2019tgz, Behera:2023nwj, Behera:2021zhi, Giacalone:2024luz}. A few of them investigate the use of hydrodynamical models~\cite{Lim:2018huo, Summerfield:2021oex, Schenke:2020mbo}, and Glauber model calculations~\cite{Rybczynski:2019adt, Sievert:2019zjr, Huang:2019tgz}. Some intend to study the parton energy loss and jet quenching effects in O--O collisions~\cite{Huss:2020whe, Zakharov:2021uza}. Interestingly, there have been a few phenomenological studies that aim to establish the effect of the presence of an $\alpha$-cluster nucleus in the final state particle production~\cite{Bozek:2014cva, Broniowski:2013dia, Li:2020vrg, Ding:2023ibq, Wang:2021ghq, Rybczynski:2017nrx, Svetlichnyi:2023nim, Behera:2023nwj, Behera:2023oxe, Giacalone:2024ixe, Zhang:2024vkh, R:2024eni}. A few of these studies employ O--O collisions, while others employ O--Au, C--Au, O--Pb, and Ne--Pb collisions. Nevertheless, the presence of a clustered nuclear structure of $^{16}$O is reflected in the final state azimuthal anisotropy when compared to a Woods-Saxon nuclear density profile. The lattice QCD estimate of a minimum initial energy density requirement of 1 GeV/$\rm fm^3$ to form a deconfined partonic medium is conjectured to be fulfilled in O--O collisions at $\sqrt{s_{\rm NN}} = 7$~TeV across all centrality classes as shown in Ref.~\cite{Behera:2021zhi}. In Ref.~\cite{Behera:2023oxe}, using a multi-phase transport model (AMPT)~\cite{Lin:2004en}, authors have shown that an away side broadening is observed in the two-particle correlation function for an $\alpha$-clustered nucleus. In addition, an increased value of $v_{3}$, and $v_{3}/v_2$ for most central O--O collisions are also reported for an $\alpha$-clustered structure. A similar observation was reported in Ref.~\cite{Bozek:2014cva}, where authors study the anisotropic flow in C--Au collisions with three $\alpha$-clusters of $^{12}$C nucleus. However, so far, no studies have shown the effect of change in nuclear density profile on event-by-event flow fluctuations. In this paper, for the first time, we report the behavior of azimuthal anisotropy and flow fluctuations in O--O collisions at $\sqrt{s_{\rm NN}}=7$ TeV using a hydrodynamically expanding system. For this study, we use a hybrid of IPGlasma initial condition, MUSIC hydrodynamics, iSS as a particlisation, and UrQMD as an afterburner for the hadronic cascade. Here, we compare results from an $\alpha$-clustered nuclear distribution with a Woods-Saxon nuclear distribution.

This paper is organized as follows. We start with a brief introduction and motivation to this study in Section~\ref{intro}. We briefly discuss the event generation and methodology in Section~\ref{sec2}. Then, we present our findings and discuss the results in Section~\ref{sec3}. Finally, we summarise our findings with a brief outlook in Section~\ref{sec4}.

\section{Event Generation and Methodology}
\label{sec2}
In this section, we briefly introduce the hybrid simulation framework used to generate events, including a discussion on the multi-particle cumulant method for the anisotropic flow estimation.

\subsection{Simulation framework}
In this work, the simulation framework implements the impact-parameter-dependent Glasma (IP-Glasma) initial conditions, MUSIC relativistic viscous hydrodynamics, particle sampling using the iSS package, and performs hadronic cascade through the UrQMD transport model. This hybrid model provides a good description of the particle production and flow in heavy-ion collisions~\cite{McDonald:2016vlt}. The key aspects of these models are also discussed briefly below, including the parameters and settings necessary to reproduce the results reported in this study.

\subsubsection{IP-Glasma: Initial condition}
\begin{figure}
\begin{center}
\includegraphics[scale=0.3]{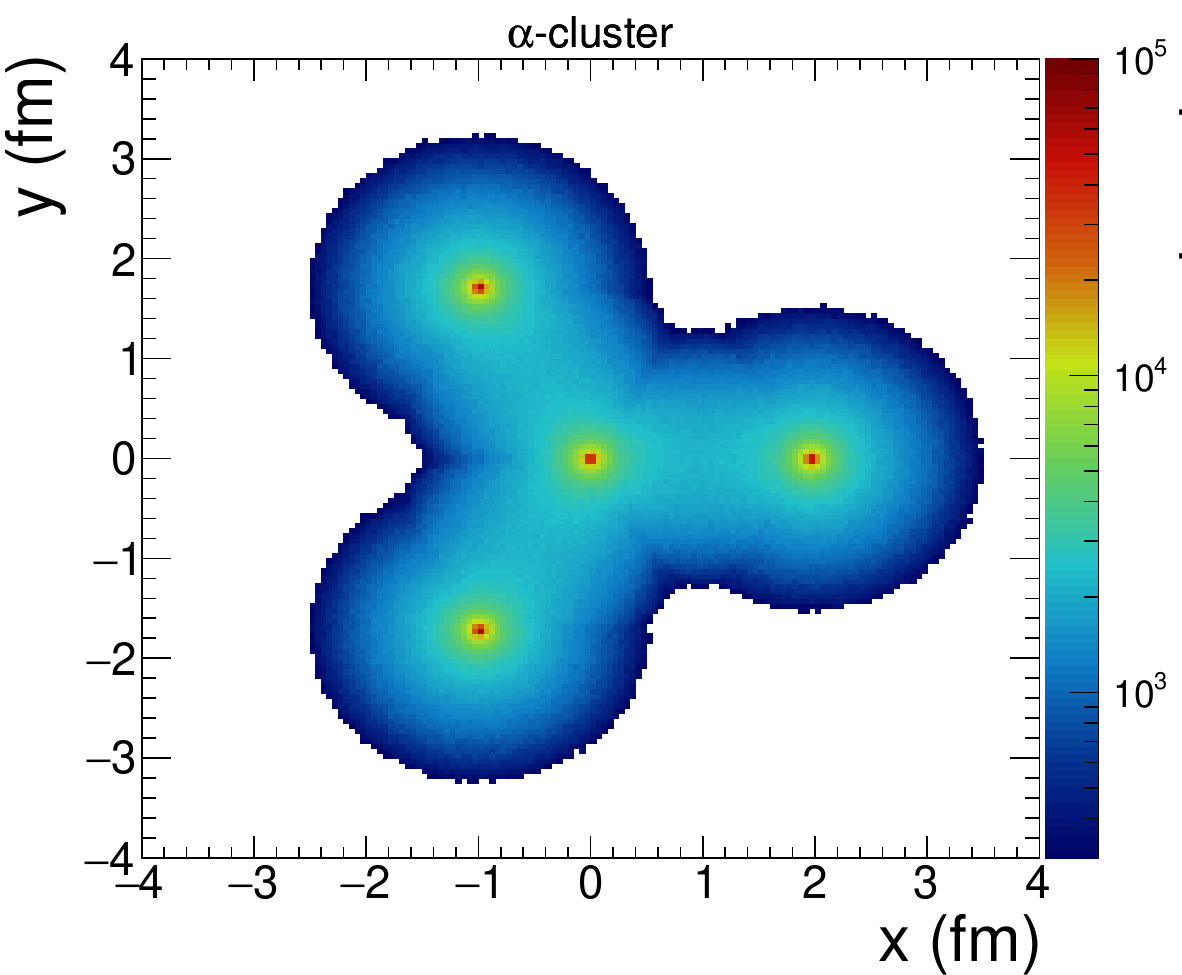}
\includegraphics[scale=0.3]{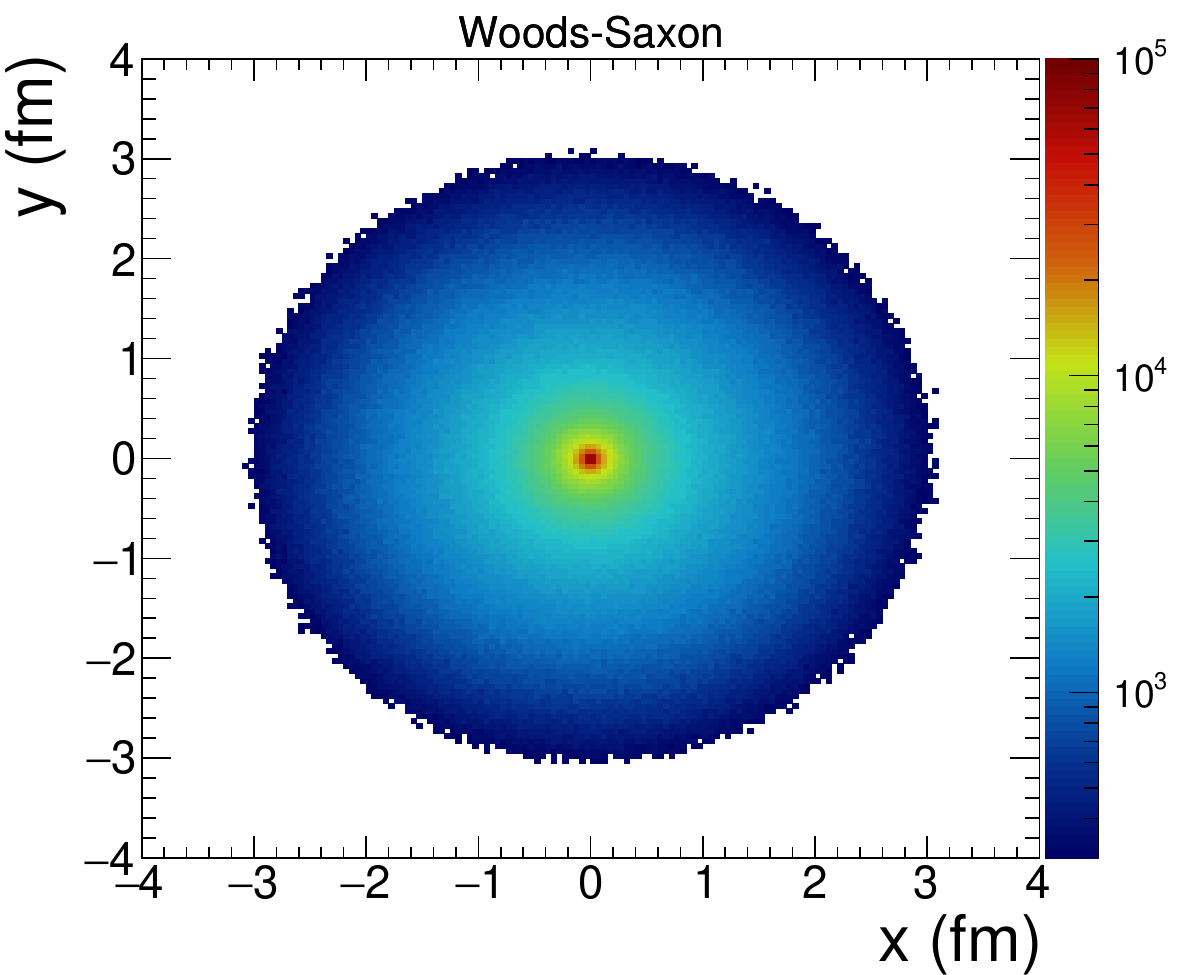}
\caption{The nucleon distribution of $^{16}$O in the $xy$ plane following $\alpha$-cluster (upper) and Woods-Saxon (lower) nuclear density profiles plotted with one million nuclei in each case.}
\label{fig:xynucldist}
\end{center}
\end{figure}

The IP-Glasma model is used to describe the dynamics of the gluon fields during the collisions of nuclear matter at relativistic speeds~\cite{Schenke:2012wb, Schenke:2012hg}. It provides the initial energy-momentum tensor of the classical Yang-Mills (CYM) color Glasma fields ($T_{\rm YM}^{\mu\nu}$), which can be further evolved using relativistic hydrodynamics. This model is based on the color glass condensate effective field theory~\cite{McLerran:1993ni, McLerran:1993ka} and uses a classical description of gluon production. This model includes fluctuations in the distributions of nucleons in the nuclear wave functions and also fluctuations in the color charge distributions inside the nucleons. The color charge fluctuations are modeled via the impact-parameter-dependent dipole saturation model (IP-Sat)~\cite{Bartels:2002cj, Kowalski:2003hm}. For the sub-nucleonic fluctuations, three hot spots are introduced in the proton thickness function, and each hot spot is parametrized using a Gaussian distribution in the transverse plane. The spatial positions of the nucleons inside the colliding nuclei are obtained from the Woods-Saxon distribution function as given below, 
\begin{equation}
\rho(r) = \rho_{0} \frac{1+ w \left(\frac{r}{r_{0}}\right)^{2}}{1 + \exp\left(\frac{r - r_{0}}{a}\right)} \ . 
\label{eq:WS}
\end{equation}
Here, $\rho(r)$ is the nuclear charge density at a radial distance $r$ from the center of the nucleus. $r_0$, $w$, and $a$ are the mean radius of the nucleus, the nuclear deformation parameter, and nuclear skin depth, respectively.
For $^{16}$O nucleus, $r_0 = 2.608$~fm, $a = 0.513$~fm, $w = -0.051$~\cite{Loizides:2014vua}. In addition, the exotic $\alpha$--cluster geometry for the $^{16}$O nucleus has also been implemented in this study. Four $\alpha$--particles ($2p,~2n$) are placed at the vertices of a regular tetrahedron with side length 3.42~fm, which makes the root mean square (RMS) radius of $^{16}$O nucleus 2.699~fm~\cite{Behera:2023nwj, Behera:2021zhi, Li:2020vrg}. The spatial positions of the nucleons inside an $\alpha$--particle are sampled using the Woods-Saxon distribution given in Eq.~\eqref{eq:WS}, with the parameters chosen for the $^4$He nucleus, which are $r_{0} = 0.964$~fm, $w = 0.517$ and $a = 0.322$~fm. This gives the RMS radius of 1.676~fm for the $\alpha$--particle. A minimum separation distance of 0.4~fm between the nucleons is imposed. Finally, to randomize each nucleus, the entire tetrahedral configuration is rotated through its center randomly in the polar and azimuthal coordinates before its initialization in IP-Glasma. Figure~\ref{fig:xynucldist} shows the nucleon distribution of $^{16}$O nuclei in the x-y plane following $\alpha$-cluster (left) and Woods-Saxon (right) nuclear density profiles. In Fig.~\ref{fig:xynucldist}, for the $\alpha$-cluster nuclear density profile of $^{16}$O nuclei, the tetrahedral configuration is not rotated for better visualization of the clustered geometry.

Once the nucleon centers are obtained, the color charges are sampled using the IP-Sat/McLerran-Venugopalan model~\cite{McLerran:1993ka, McLerran:1994vd}. Now, color currents arise due to the motion of color charges present inside the colliding nuclei along the beam direction. These currents act as the sources in the Yang-Mills equation,
$[D_\mu, F^{\mu\nu}] = J^\nu$. The initial condition at $\tau=0$ is given by the solution of CYM equations in the Schwinger gauge. For the subsequent evolution of the equations of motion for Glasma fields, a lattice formulation has been used, which was first introduced in Ref.~\cite{Krasnitz:1998ns}. These equations of motion are solved using a leapfrog algorithm on a 2D lattice. For our case, we use the lattice size, $L=14$, and lattice spacing $a=0.02$~fm~\cite{Schenke:2020mbo}. Finally, all the component of the energy-momentum tensor ($T_{\rm YM}^{\mu\nu}$) for the Yang-Mills system is computed at $\tau_{\rm switch}=0.4$~fm, and the initial-state configuration is transferred to the hydrodynamic simulation.

\subsubsection{MUSIC: Hydrodynamics}
Under the assumption of system under local thermal equilibrium, the initial energy-momentum tensor obtained from the IP-Glasma model at thermalization time $\tau_0=0.4$~fm is evolved using the MUSIC framework of relativistic hydrodynamics following the conservation law $\partial_\mu T^{\mu\nu} = 0$~\cite{Schenke:2010nt, Schenke:2010rr, Paquet:2015lta}. The energy-momentum tensor is composed of both ideal and dissipative parts, which are combined to give $T^{\mu\nu} = eu^{\mu}u^{\nu}-(P+\Pi)\Delta^{\mu\nu}+\pi^{\mu\nu}$. Here, $e$ is the energy density, $u^{\mu}$ is the fluid flow four-velocity, $P$ is the thermodynamic pressure, $\Pi$ is the bulk viscous pressure, and $\pi^{\mu\nu}$ is the shear viscous tensor. The projection operator is defined as $\Delta^{\mu\nu} = g^{\mu\nu}-u^{\mu}u^{\nu}$. In this study, we use the boost invariant picture of hydrodynamics and, hence, perform a (2+1)-dimensional evolution. The equation-of-state (EoS) parametrization \texttt{"s95p-v1.2"} is used to close the system of equations in hydrodynamics. This EoS is obtained from the interface of lattice data and hadron resonance gas~\cite{Huovinen:2009yb}. We use a fixed $\eta/s = 0.12$, and a temperature dependent $\zeta/s(T)$ as described in Ref.~\cite{Schenke:2020mbo}. In MUSIC, the hydrodynamic equations are solved using the Kurganov-Tadmor (KT) algorithm~\cite{Kurganov:2000ovy, Jeon:2015dfa}.

\subsubsection{iSS: Particle sampler}
The fluid-dynamic description is finally switched to the particle description when the local energy density drops to the switching energy density, $e_{\rm sw} = 0.18$~GeV/fm$^3$~\cite{Schenke:2020mbo}. At this point, the freeze-out hypersurface from MUSIC is fed to the iSS particle sampler~\cite{Shen:2014vra, Denicol:2018wdp}, which follows the usual Cooper-Frye formalism for computing the particle spectra using equilibrium distribution with both shear and bulk viscous correction terms~\cite{Cooper:1974mv, Dusling:2009df, Schenke:2020mbo}. Each MUSIC freeze-out hypersurface can be over-sampled using the iSS particle sampler to increase event statistics and save computation time. So, we have sampled 200 events from each MUSIC output hypersurface.

\subsubsection{UrQMD: Hadronic cascade}
Finally, the produced particles are evolved using the UrQMD microscopic transport model~\cite{Bass:1998ca, Bleicher:1999xi}. This model handles both elastic and inelastic processes of hadronic scatterings and resonance decay with particles of masses up to 2.25 GeV. UrQMD closely simulates a realistic description of the final-stage hadron evolution and performs a dynamical freeze-out for different particle species. All parameters in the UrQMD model are set to their default values. The output of UrQMD includes final-state particle four momenta and particle identification (PID) code which can be written to disk for further analysis.

\subsection{Multi-particle cumulant method}

Anisotropic flow can be expressed as the coefficients of Fourier expansion, as shown in Eq.~\eqref{eq:fourierexpansion}. One can estimate the anisotropic flow coefficients using the following equation~\cite{Mallick:2023vgi, Mallick:2022alr, Prasad:2022zbr}.
\begin{equation}
    v_{n}=\langle \cos[n(\phi-\psi_{n})] \rangle \
\label{eq:eventplane}
\end{equation}
Here, $\langle \dots \rangle$ stands for the average over all the charged particles in a single event. Obtaining the anisotropic flow coefficients using Eq.~\eqref{eq:eventplane} requires $\psi_{n}$, whose estimation is not trivial in experiments. 
To avoid the requirement to estimate the symmetry plane angle, we use a multi-particle Q-cumulant method~\cite{Bilandzic:2010jr, Bilandzic:2013kga}.
Since we aim to estimate the anisotropic flow fluctuations, thus for this study, we limit our measurements to two- and four-particle cumulants, which can be estimated using the flow vector ($Q_{n}$) defined as follows.
\begin{equation}
Q_{n} = \sum_{j=1}^{M}e^{in\phi_{j}},
\end{equation}
where $M$ is the multiplicity of the event, and $\phi$ is the azimuthal angle of the charged hadrons. One can obtain a single event average two- and four-particle correlation function using the following expressions.
\begin{equation}
\begin{aligned}
\langle 2 \rangle & = \frac{|Q_{n}|^{2} - M} {M(M-1)}, \\
\langle 4 \rangle & = \frac{|Q_{n}|^{4} + |Q_{2n}|^{2} - 2 \cdot {\rm{Re}}[Q_{2n}Q_{n}^{*}Q_{n}^{*}]  } { M(M-1)(M-2)(M-3) }  \\
& ~ ~ - 2 \frac{ 2(M-2) \cdot |Q_{n}|^{2} - M(M-3) } { M(M-1)(M-2)(M-3)},
\end{aligned}
\label{Eq:Mean24}
\end{equation}
One can obtain the event-average correlation function using the following expression.
\begin{equation}
\begin{aligned}
    \langle \langle 2 \rangle \rangle & = \frac{\sum_{i=1}^{N_{\rm ev}}(W_{\langle 2 \rangle})_{i}\langle 2 \rangle_{i}}{\sum_{i=1}^{N_{\rm ev}}(W_{\langle 2 \rangle})_{i}}, \\
    \langle \langle 4 \rangle \rangle & = \frac{\sum_{i=1}^{N_{\rm ev}}(W_{\langle 4 \rangle})_{i}\langle 4 \rangle_{i}}{\sum_{i=1}^{N_{\rm ev}}(W_{\langle 4 \rangle})_{i}}.
\end{aligned}
\end{equation}
Here, $\langle\langle \dots \rangle\rangle$ denotes the average taken over all the charged hadrons over all events. The `*' stands for the complex conjugate. $N_{\rm ev}$ is the total number of events used for the measurements. Finally, $(W_{\langle 2 \rangle})_{i}$ and $(W_{\langle 4 \rangle})_{i}$ are the weight factors for the $i$\textsuperscript{th} event which takes into account the number of different two- and four-particle combinations in the event of multiplicity $M$. Weight factors, $W_{\langle 2 \rangle}$ and $W_{\langle 4 \rangle}$ can be calculated using the following equations.
\begin{equation}
\begin{aligned}
        W_{\langle 2 \rangle} & = M(M-1), \\
        W_{\langle 4 \rangle} & = M(M-1)(M-2)(M-3).
\end{aligned}
\end{equation}
Consequently, the two- and four-particle cumulants can be obtained as:
\begin{equation}
\begin{aligned}
c_{n}\{2\} & = \langle \langle 2 \rangle \rangle, \\
c_{n}\{4\} & = \langle \langle 4 \rangle \rangle - 2 \cdot \langle \langle 2 \rangle \rangle^{2},
\end{aligned}
\label{Eq:c24}
\end{equation}
Now, the reference flow of the particles can be estimated using the following expression.
\begin{equation}
\begin{aligned}
    v_{n}\{2\} & = \sqrt{c_{n}\{2\}}, \\
    v_{n}\{4\} & = \sqrt[4]{-c_{n}\{4\}}.
\end{aligned}
\label{eq:v24}
\end{equation}
To estimate the differential flow of the particles of interest (POIs), the $p_{n}$ and $q_{n}$ vectors for specific kinematic range and/or for specific hadron species are defined as follows:
\begin{equation}
\begin{aligned}
p_{n}& = \sum_{j=1}^{m_{p}} e^{in\phi_{j}},\\
q_{n}& = \sum_{j=1}^{m_{q}} e^{in\phi_{j}},
\label{Eqpvector}
\end{aligned}
\end{equation}
where $m_{p}$ is the total number of particles labeled as POIs, and $m_{q}$ is the total number of particles tagged as both reference flow particles (RFPs) and POIs. RFPs serve as a reference frame for POIs to quantify the collective motion of the system and help establish the orientation of the event plane, which is crucial for determining the anisotropic flow coefficients. The single-event average differential two- and four-particle azimuthal correlation functions are calculated as follows.
\begin{equation}
\begin{aligned}
\langle 2^{'} \rangle &= \frac{p_{n} Q_{n}^{*} - m_{q}} {m_{p}M - m_{q}},  \\
\langle 4^{'} \rangle  & = \left[p_{n}Q_{n}Q_{n}^{*}Q_{n}^{*}  - q_{2n}Q_{n}^{*}Q_{n}^{*} - p_{n}Q_{n}Q_{2n}^{*} \right. \\
& - 2\cdot M p_{n} Q_{n}^{*} - 2 \cdot m_{q} |Q_{n}|^{2}  + 7 \cdot q_{n}Q_{n}^{*} - Q_{n}q_{n}^{*} \\
& \left. + q_{2n}Q_{2n}^{*} + 2 \cdot p_{n} Q_{n}^{*} + 2 \cdot m_{q}M - 6 \cdot m_{q} \right] \\
&  / \left[ (m_{p}M - 3 m_{q})(M-1) (M-2) \right]
\end{aligned}
\label{Eq:Mean24p}
\end{equation}

Similarly, one obtains the event-average differential azimuthal correlations using the following expressions.

\begin{equation}
\begin{aligned}
        \langle\langle 2^{'} \rangle\rangle & = \frac{\sum_{i=1}^{N_{\rm ev}}(w_{\langle 2^{'} \rangle})_{i}\langle 2^{'} \rangle_{i}}{\sum_{i=1}^{N_{\rm ev}}(w_{\langle 2^{'} \rangle})_{i}}, \\
        \langle\langle 4^{'} \rangle\rangle & = \frac{\sum_{i=1}^{N_{\rm ev}}(w_{\langle 4^{'} \rangle})_{i}\langle 4^{'} \rangle_{i}}{\sum_{i=1}^{N_{\rm ev}}(w_{\langle 4^{'} \rangle})_{i}}.
\end{aligned}
\end{equation}
Here, $w_{\langle 2^{'} \rangle}$ and $w_{\langle 4^{'} \rangle}$ are the weights corresponding to two- and four-particle cumulants, given by:
\begin{equation}
\begin{aligned}
    w_{\langle 2^{'} \rangle} & = m_{p}M - m_{q}, \\
    w_{\langle 4^{'} \rangle} & = (m_{p}M - 3 m_{q})(M-1) (M-2) .
\end{aligned}
\end{equation}
Thus, one finds the two- and four-particle differential cumulants using the following equations:
\begin{equation}
\begin{aligned}
d_{n}\{2\} & = \langle \langle 2^{'} \rangle \rangle, \\
d_{n}\{4\} & = \langle \langle 4^{'} \rangle \rangle  - 2 \langle \langle 2^{'}\rangle \rangle \langle \langle 2\rangle \rangle.
\label{Eqdn24}
\end{aligned}
\end{equation}

Finally, one can calculate the differential flow $v_{2}(p_{\rm T})$ using  two- and four-particle correlations as follows:
\begin{equation}
\begin{aligned}
v_{n}\{2\}(p_{\rm T}) & = \frac{d_{n}\{2\}}{\sqrt{c_{n}\{2\}}},  \\
v_{n}\{4\}(p_{\rm T}) & = -\frac{d_{n}\{4\}}{\sqrt[4]{(-c_{n}\{4\})^3}}.
\end{aligned}
\label{eq:vncnrelation}
\end{equation}

\begin{figure}
\begin{center}
\includegraphics[scale=0.25]{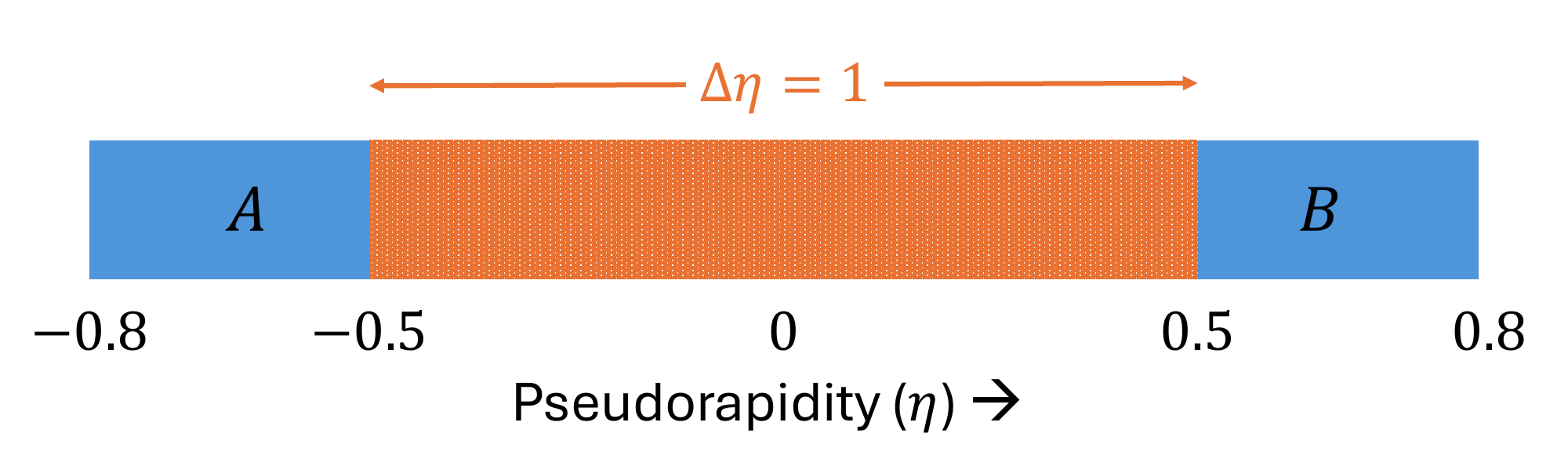}
\caption{Pictorial representation of the pseudorapidity regions $A$ and $B$ separated with a pseudorapidity gap ($\Delta\eta$).}
\label{fig:deltaeta}
\end{center}
\end{figure}

Unfortunately, the $v_{n}$ obtained from the two-particle Q-Cumulant method contains contributions from non-flow effects, which can be suppressed by appropriate kinematic cuts. One can introduce a pseudorapidity gap between the particles in the two-particle Q-cumulant method~\cite{ALICE:2011ab, Aamodt:2010pa, Zhou:2014bba, Zhou:2015iba, ALICE:2014dwt} to suppress the non-flow contributions, which is equivalent to the scalar product method of calculating the anisotropic flow coefficients~\cite{STAR:2002hbo}. Accordingly, the whole event is divided into two sub-events~\cite{Jia:2017hbm, ATLAS:2017rtr, ATLAS:2017hap}, $A$ and $B$, which are separated by a $|\Delta\eta|$ gap, as depicted in Fig.~\ref{fig:deltaeta}. This modifies Eq.~\eqref{Eq:Mean24} to:
\begin{equation}
\langle 2 \rangle _{\Delta \eta} = \frac{Q_{n}^{A} \cdot Q_{n}^{B *}} {M_{A} \cdot M_{B}},
\label{Eq:Mean2Gap}
\end{equation}
where $Q_{n}^{A}$ and $Q_{n}^{B}$ are the flow vectors from the sub-event $A$ and $B$, respectively. $M_{A}$ and $M_{B}$ are the corresponding multiplicities. Thus, the two-particle Q-cumulant with a $|\Delta\eta|$ gap is given by:
\begin{equation}
c_{n}\{2, |\Delta\eta|\} = \langle \langle 2 \rangle \rangle _{\Delta\eta} .
\label{Eq:v22Gap10}
\end{equation}
In calculations of differential flow with a pseudorapidity gap, there is no overlap of POIs and RFPs if we select RPs from one subevent and POIs from the other. Thus Eq.~\eqref{Eq:Mean24p} can be modified to:
\begin{equation}
\langle 2^{'} \rangle_{\Delta \eta}  = \frac{p_{n,A} Q_{n,B}^{*} } {m_{p,A}M_{B}},
\label{Eqmean2pGap}
\end{equation}
and we obtain the differential two-particle cumulant as:
\begin{equation}
d_{n}\{2, |\Delta\eta|\}  = \langle \langle 2^{'} \rangle \rangle_{\Delta\eta}.
\label{dn2Gap}
\end{equation}
Finally, the differential flow from the two-particle cumulant can be obtained by inserting the two-particle reference flow (with $\eta$ gap) into the differential two-particle cumulant:
\begin{equation}
v_{n}\{2, |\Delta\eta| \}(p_{\rm T})  = \frac{d_{n}\{2, |\Delta\eta|\}}{\sqrt{c_{n}\{2,|\Delta\eta|\}}}.
\label{vn2EtaGap}
\end{equation}
In this paper, the elliptic and triangular flow coefficients are calculated using the above equations by setting $n=$ 2 and 3, respectively. Considering small and Gaussian-like fluctuations of anisotropic flow coefficients ($\sigma_{v_{n}}$), the mean ($\langle v_{n}\rangle$) and fluctuations of anisotropic flow coefficients can be represented in terms of $v_{n}\{2, |\Delta\eta|\}$ and $v_{n}\{4\}$, as follows~\cite{PHENIX:2018lfu, Ollitrault:2009ie}.
\begin{equation}
    \langle v_{n}\rangle=\sqrt{\frac{v_{n}^{2}\{2, |\Delta\eta|\}+v_{n}^{2}\{4\}}{2}}
    \label{eq:meanv2}
\end{equation}
\begin{equation}
    \sigma_{v_{n}}=\sqrt{\frac{v_{n}^{2}\{2, |\Delta\eta|\}-v_{n}^{2}\{4\}}{2}}
    \label{eq:sigmav2}
\end{equation}

Using Eqs. \eqref{eq:meanv2}, and \eqref{eq:sigmav2}, one can obtain the relative anisotropic flow fluctuation ($F(v_{n})$) as follows.

\begin{equation}
    F(v_{n})=\frac{\sigma_{v_{n}}}{\langle v_{n}\rangle}
\end{equation}

In this study, we use 3500 IPGlasma + MUSIC events, and each IPGlasma + MUSIC event is sampled 200 times for hadron rescattering using UrQMD. Thus, we have a total of 700K minimum bias events. We use all the charged hadrons within the pseudorapidity region, $|\eta|<0.8$, for the estimation of anisotropic flow coefficients using the two- and four-particle Q-cumulants method. We tag the charged hadrons within $|\eta|<0.8$ and $0.2<p_{\rm T}<4.0$ GeV/c as RFPs. We apply a pseudorapidity gap, $|\Delta\eta|>1.0$ in the two-subevent method, as shown in Fig.~\ref{fig:deltaeta}, to subtract the non-flow effects from two-particle Q-cumulants method. The statistical uncertainties are calculated using the relations shown in Ref.~\cite{Bilandzic:2012wva}.

\section{Results and discussions}
\label{sec3}

In this section, we start with a discussion of the multiplicity selection, followed by discussions on the results for anisotropic flow coefficients. Finally, we discuss the multiplicity dependence of flow fluctuations.

\begin{figure}
\begin{center}
\includegraphics[scale=0.4]{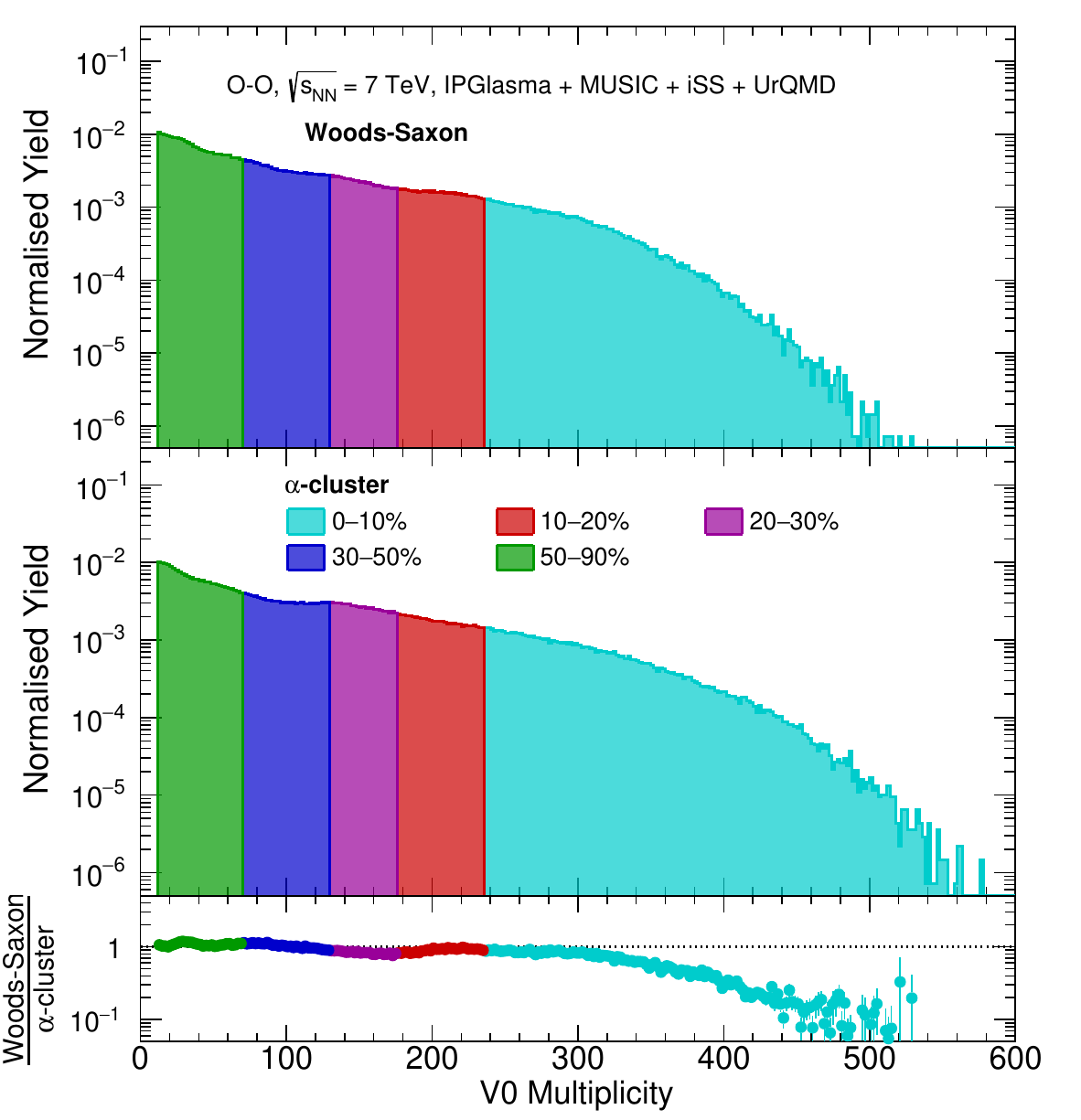}
\caption{Distribution of charged particle produced in the pseudorapidity acceptance of the V0 detector of ALICE, i.e., $-3.7<\eta<-1.7$ and $2.8<\eta<5.1$ (V0 Multiplicity or V0M) in O--O collisions at $\sqrt{s_{\rm NN}}=7$ TeV using IPGlasma + MUSIC + iSS + UrQMD model for Woods-Saxon (upper) and $\alpha$-cluster (middle) nuclear density profiles. The lower panel shows the ratio of V0M distributions from Woods-Saxon to that of in $\alpha$-cluster nuclear profile.}
\label{fig:V0Mdist}
\end{center}
\end{figure}

Figure~\ref{fig:V0Mdist} shows the distribution of charged particles produced in the pseudorapidity region $-3.7<\eta<-1.7$ and $2.8<\eta<5.1$ (V0M) in O--O collisions at $\sqrt{s_{\rm NN}}=7$ TeV using IPGlasma + MUSIC + iSS + UrQMD model for the Woods-Saxon (upper) and $\alpha$-cluster (lower) nuclear density profiles. The lower panel shows the ratio of normalized yields of V0M for Woods-Saxon and $\alpha$-cluster nuclear distributions. Also, different multiplicity classes are defined based on the percentile slices on the V0M distribution for the Woods-Saxon nuclear density profile. Here, for a fair comparison, we use the same cuts of V0M distribution for all the multiplicity classes for both the nuclear density profiles. The first multiplicity class, e.g., 0--10\%, has the highest particle multiplicity, while the last multiplicity class, 50--90\%, has the events having the lowest final state charged particle multiplicity, as shown in Fig.~\ref{fig:V0Mdist}. Thus, one may infer that the 0--10\% class refers to the most central collisions, while the last multiplicity class, 50--90\% in this study, implies a peripheral collision. Further, one can refer to Fig.~\ref{fig:ImpBV0M} in the Appendix~\ref{app:impB} for the average value of the impact parameter corresponding to each multiplicity class for both Woods-Saxon and $\alpha$-cluster nuclear density profiles. In Fig.~\ref{fig:V0Mdist}, one finds a higher yield for V0M for the O--O collisions having an $\alpha$-clustered nucleus. In addition, for a similar value of V0M, one finds a higher yield for the $\alpha$-clustered distribution of nucleons inside the $^{16}$O nuclei for V0M $>300$, which can be seen in the bottom ratio plot. In Ref.~\cite{Behera:2021zhi}, the probability distribution of $\alpha$-clustered nuclear density profile of $^{16}$O nuclei is compared with Woods-Saxon and Harmonic-Oscillator nuclear density profiles. Here, one finds that the probability of finding nucleons in Harmonic-Oscillator and Woods-Saxon nuclear density profiles spreads over long distances from the center of $^{16}$O nuclei. In contrast, the probability of finding nucleons in the $\alpha$-cluster profile is limited to some finite distance from the center of the nucleus. This demonstrates the compactness of the $\alpha$-clustered structure of $^{16}$O nuclei, which may lead to increased energy density, as shown in Ref.~\cite{Behera:2021zhi}. A large energy density can lead to an increase in final state particle yield, as can be seen in Fig.~\ref{fig:V0Mdist}.


\begin{figure}
\begin{center}
\includegraphics[scale=0.4]{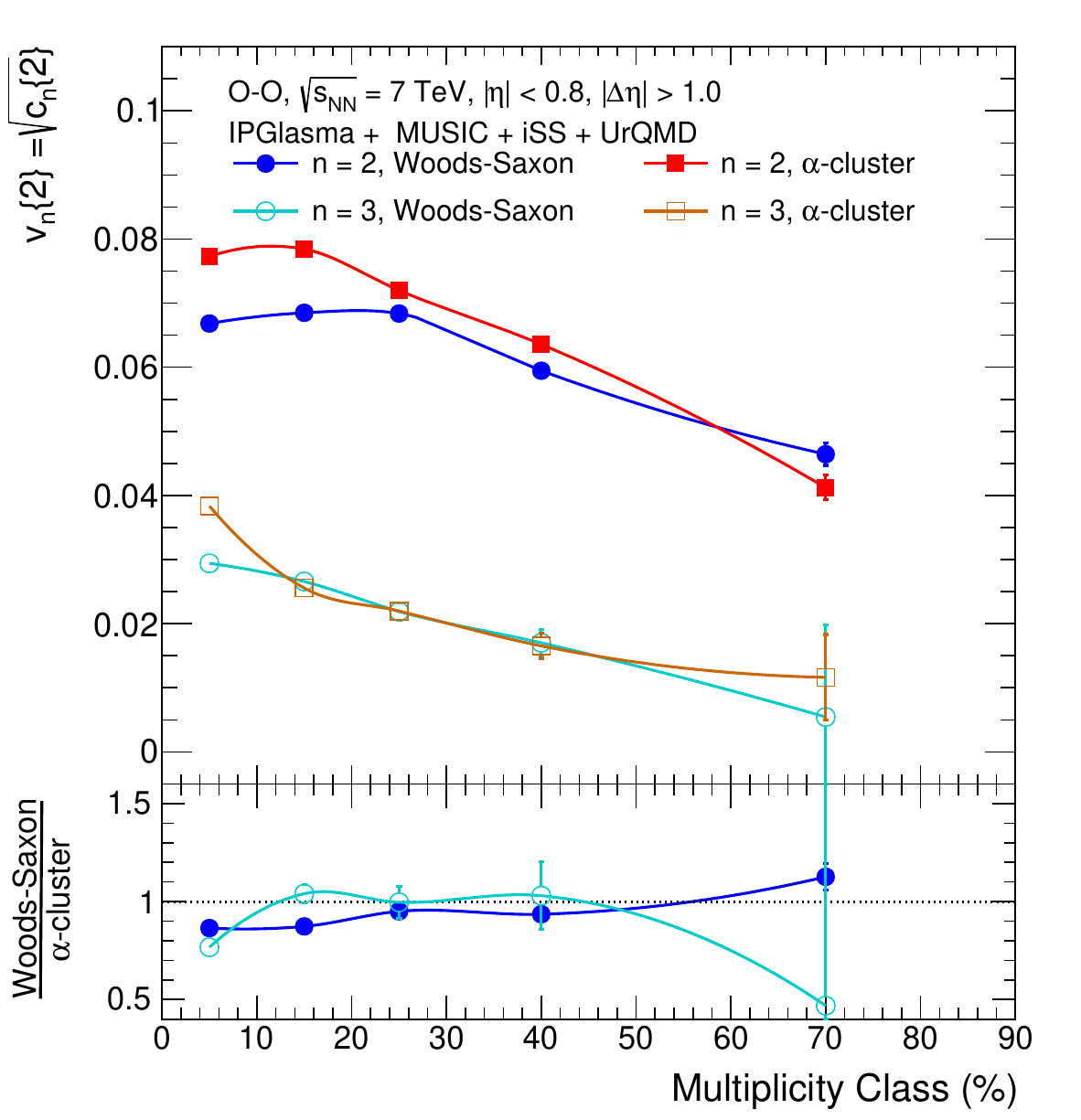}
\caption{Upper panel shows the elliptic flow and triangular flow using two-particle Q-cumulant method as a function of multiplicity class (\%) in O--O collisions at $\sqrt{s_{\rm NN}}=7$ TeV using IPGlasma + MUSIC + iSS + UrQMD model for Woods-Saxon and $\alpha$-cluster nuclear density profiles. The lower panel shows the ratio of $v_{n}\{2\}$ for Woods-Saxon to a $\alpha$-cluster nuclear density profile.}
\label{fig:vn2etagap}
\end{center}
\end{figure}

\begin{figure}
\begin{center}
\includegraphics[scale=0.4]{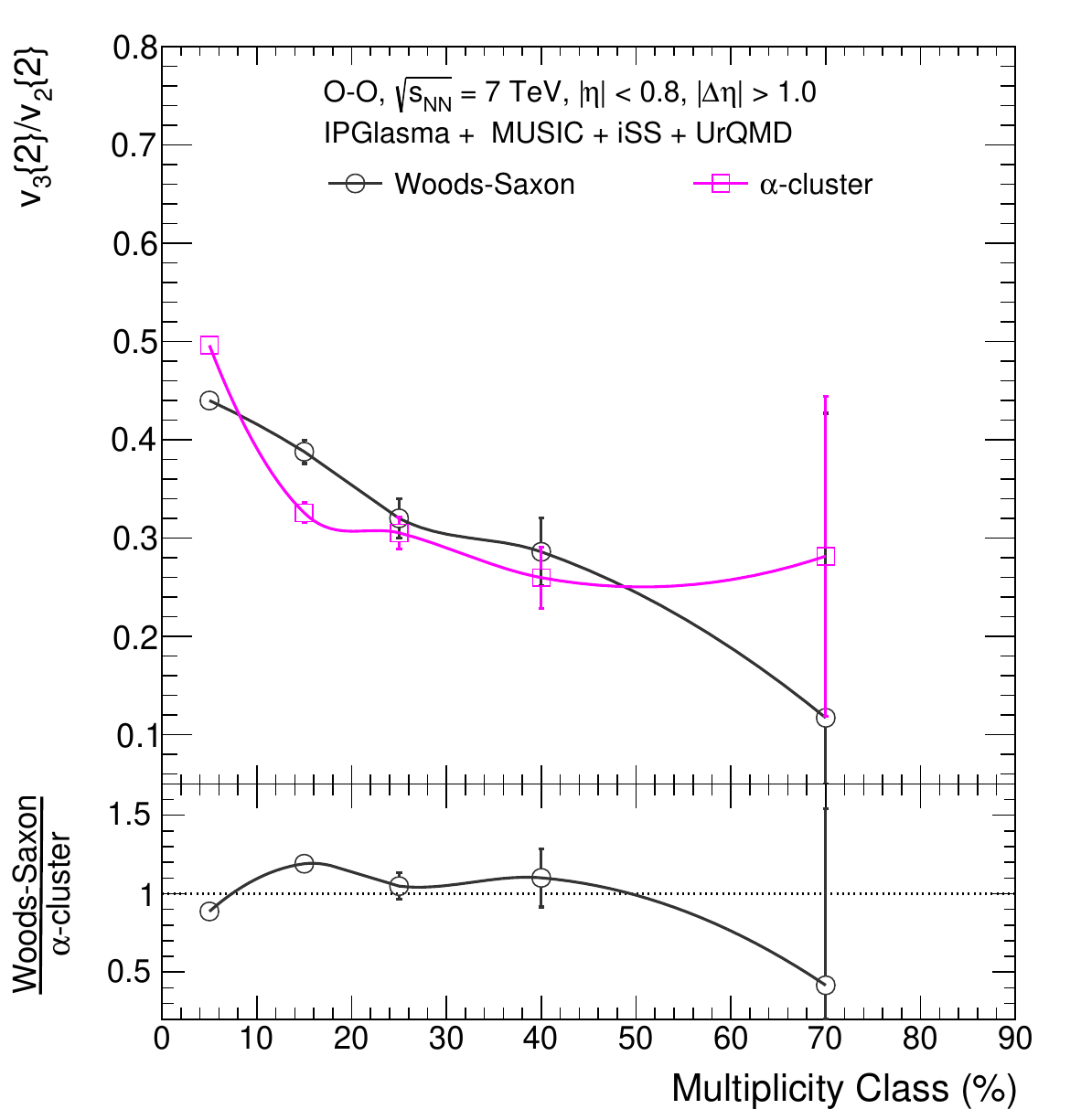}
\caption{Upper panel shows the ratio $ v_{3}\{2, |\Delta\eta|>1.0\}/ v_{2}\{2, |\Delta\eta|>1.0\}$ as a function of multiplicity class (\%) in O--O collisions at $\sqrt{s_{\rm NN}}=7$ TeV using IPGlasma + MUSIC + iSS + UrQMD model for Woods-Saxon and $\alpha$-cluster nuclear density profiles. Lower panel shows the ratio of $ v_{3}\{2, |\Delta\eta|>1.0\}/ v_{2}\{2, |\Delta\eta|>1.0\}$ from Woods-Saxon to a $\alpha$-cluster nuclear density profile.}
\label{fig:ratiov22v32}
\end{center}
\end{figure}

Figure~\ref{fig:vn2etagap} shows $v_{2}\{2, |\Delta\eta|>1.0\}$ and $v_{3}\{2, |\Delta\eta|>1.0\}$ with a pseudorapidity gap, $|\Delta\eta|>1.0$, as a function of multiplicity class for Woods-Saxon and $\alpha$-cluster nuclear density profiles of $^{16}$O in O--O collisions at $\sqrt{s_{\rm NN}}=7$ TeV using IPGlasma + MUSIC + iSS + UrQMD model. One notices a moderate multiplicity dependence on both elliptic and triangular flow in both Woods-Saxon and $\alpha$-cluster nuclear density profiles. The values of $v_{2}\{2, |\Delta\eta|>1.0\}$ for the $\alpha$-cluster nuclear density profile are higher than the Woods-Saxon nuclear density profiles throughout the multiplicity classes except for the 50--90\% case, where the trend is reversed. This nuclear density profile dependence of $v_{2}\{2, |\Delta\eta|>1.0\}$ has significantly large effects from the initial state eccentricities, as shown in Fig.~\ref{fig:ecc}.
Further, the value of elliptic flow is the largest for the 0--10\% multiplicity class and decreases as the final state particle multiplicity decreases. In the 0--10\% multiplicity class, the observed high values of elliptic flow can be attributed to the initial eccentricity fluctuations. In contrast, towards the low multiplicity regions, although the value of eccentricity increases, the continued decrease in the system size and lifetime takes over, restricting the complete transfer of initial eccentricity to the final state azimuthal anisotropy. This results in a smaller value of anisotropic flow coefficients towards the low multiplicity regions than for the high multiplicity events. However, the observed behavior of elliptic flow as a function of multiplicity class and nuclear density profile assuming a hydrodynamical expansion vary significantly from that observed in a kinetic theory-based model, such as AMPT~\cite{Behera:2023nwj}. In contrast, we observe an interesting behavior for $v_{3}\{2, |\Delta\eta|>1.0\}$ as a function of multiplicity class for both the nuclear density profiles. One finds that $v_{3}\{2, |\Delta\eta|>1.0\}$ decreases from 0--10\% to 50--90\% multiplicity classes, having similar slopes for the Woods-Saxon and $\alpha$-clustered nuclear density profiles within 20--90\% multiplicity classes. This consistency in the triangular flow between two nuclear density profiles for the mentioned multiplicity class is consistent with the AMPT predictions, as shown in Ref.~\cite{Behera:2023nwj}. Interestingly, it can be observed from the lower panel of Figure~\ref{fig:vn2etagap}, a sudden rise of $v_{3}\{2, |\Delta\eta|>1.0\}$ in the 0--10\% multiplicity class for the $\alpha$-clustered in contrast to a Woods-Saxon density profile. These results are similar to the AMPT model calculations in Ref.~\cite{Behera:2023nwj}. We believe that the sudden rise in $v_{3}\{2, |\Delta\eta|>1.0\}$ for 0--10\% multiplicity class of $^{16}$O nuclei having an $\alpha$-clustered nuclear density profile can be attributed to the presence of an initial tetrahedral arrangement of four $\alpha$-particles inside the $^{16}$O nucleus.

\begin{figure*}
\begin{center}
\includegraphics[scale=0.4]{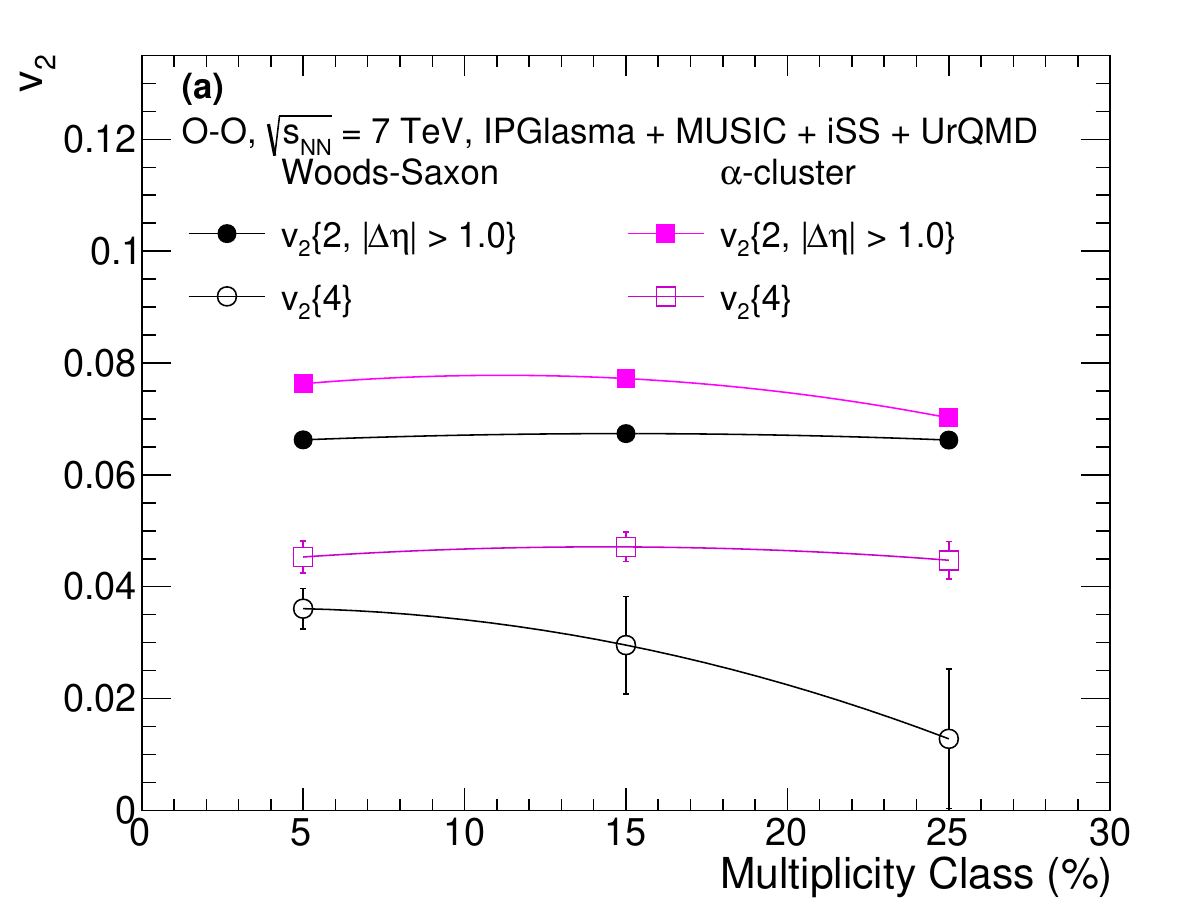}
\includegraphics[scale=0.4]{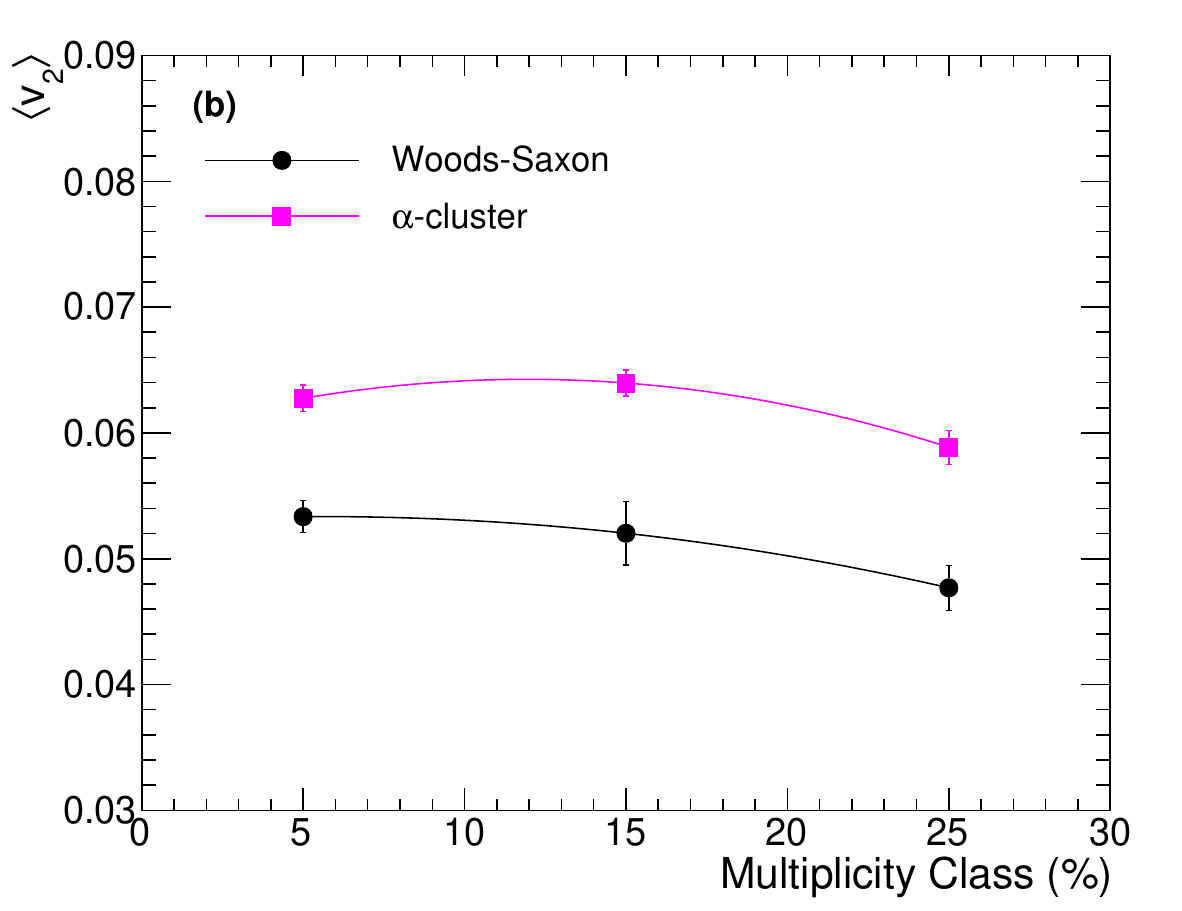}
\includegraphics[scale=0.4]{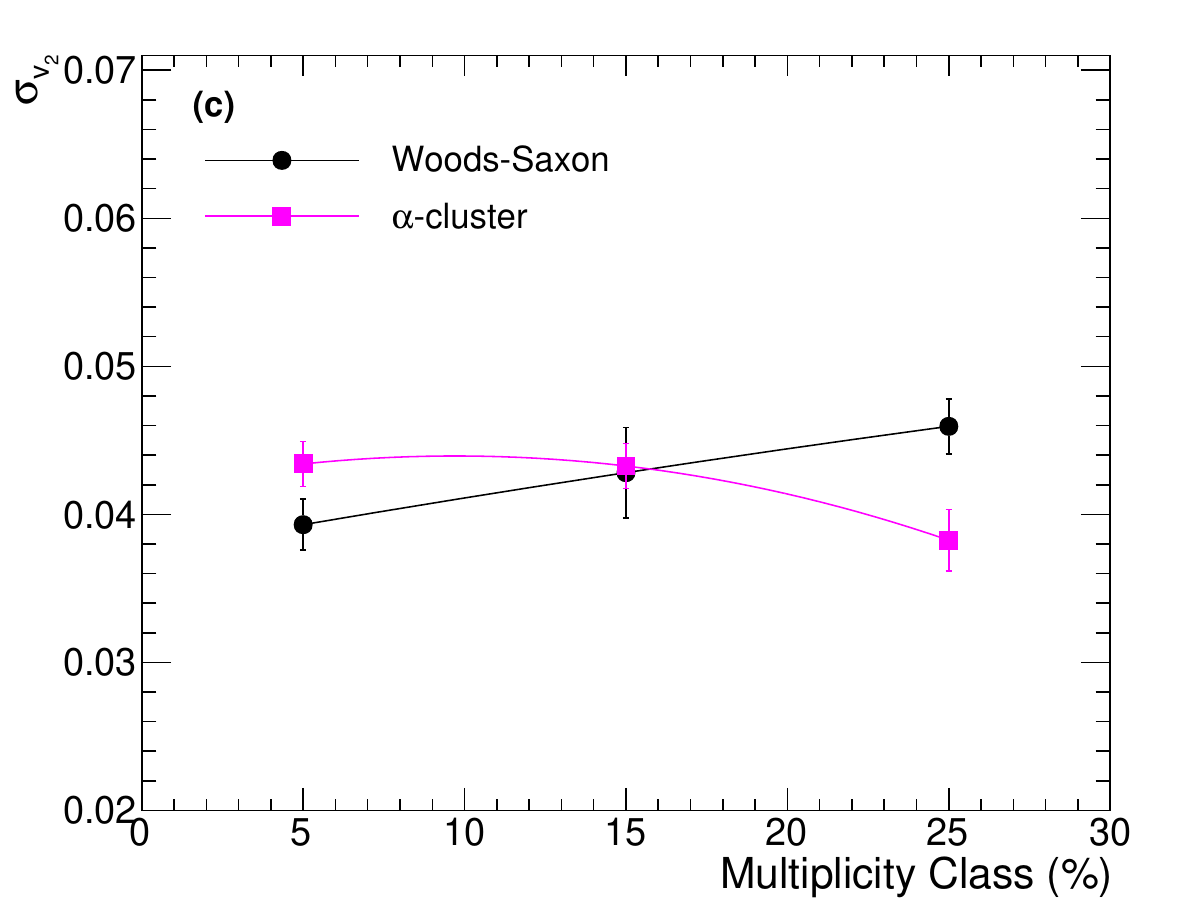}
\includegraphics[scale=0.4]{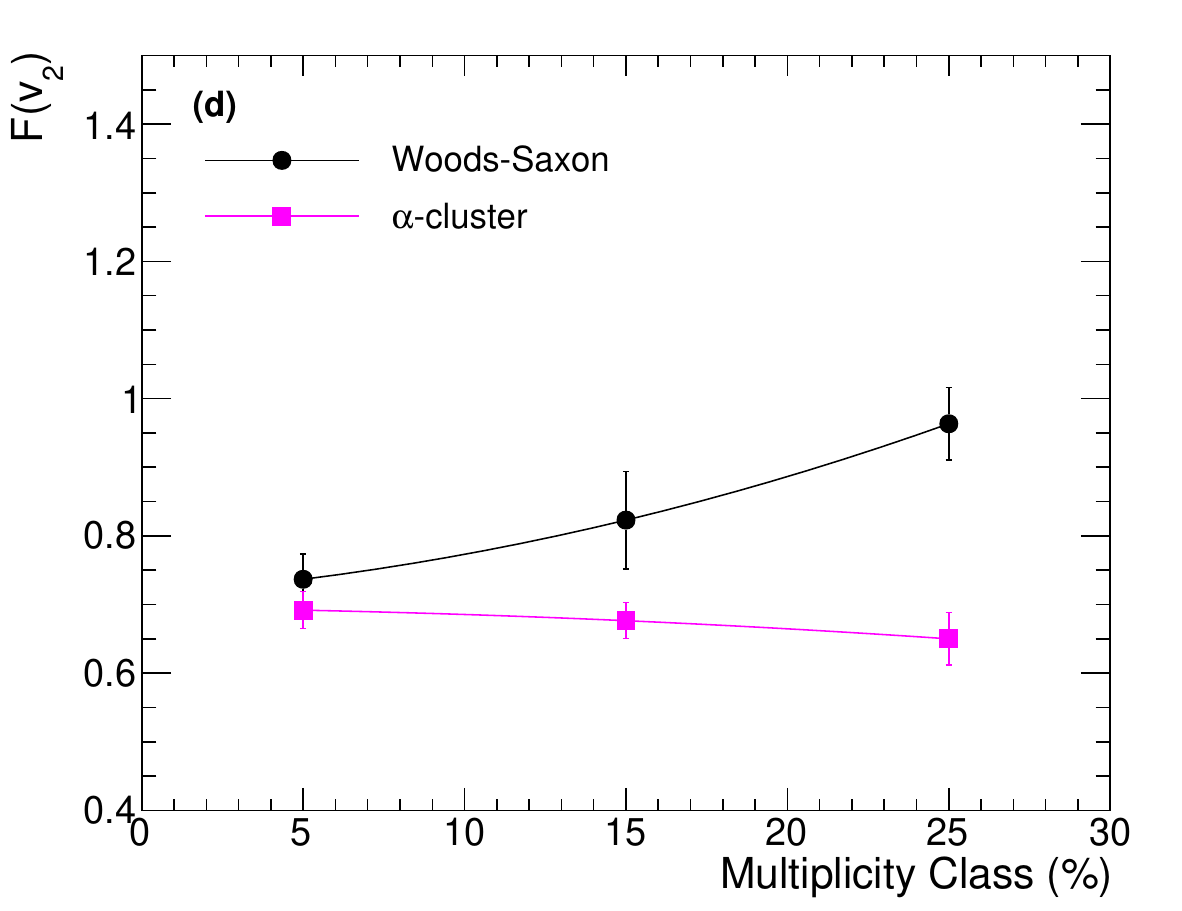}
\caption{Multiplicity class dependence of $v_{2}$, $\langle v_{2}\rangle$, $\sigma_{v_{2}}$, and $F(v_{2})$ in O--O collisions, from (a) to (d), respectively, at $\sqrt{s_{\rm NN}}=7$ TeV using IPGlasma + MUSIC + iSS + UrQMD model for Woods-Saxon and $\alpha$-cluster nuclear density profiles.}
\label{fig:v2fluctuation}
\end{center}
\end{figure*}

The upper panel of Fig.~\ref{fig:ratiov22v32} shows the ratio $ v_{3}\{2, |\Delta\eta|>1.0\}/ v_{2}\{2, |\Delta\eta|>1.0\}$ as a function of multiplicity class in O--O collisions at $\sqrt{s_{\rm NN}}=7$ TeV using IPGlasma + MUSIC + iSS + UrQMD model for Woods-Saxon and $\alpha$-cluster nuclear density profiles. The bottom panel shows the ratio of $ v_{3}\{2, |\Delta\eta|>1.0\}/ v_{2}\{2, |\Delta\eta|>1.0\}$ for Woods-Saxon to an $\alpha$-cluster nuclear density profile. Interestingly, when the $^{16}$O nuclei having the Woods-Saxon nuclear distributions collide, their elliptic and triangular flow values follow a modest dependence on final state multiplicity, as one can see in Fig.~\ref{fig:vn2etagap}. Thus, the ratio $ v_{3}\{2, |\Delta\eta|>1.0\}/ v_{2}\{2, |\Delta\eta|>1.0\}$ also exhibits a modest decrease with a decrease in particle multiplicity. However, one finds a notably stronger increase in  $ v_{3}\{2, |\Delta\eta|>1.0\}/ v_{2}\{2, |\Delta\eta|>1.0\}$ for 0--10\% multiplicity class for the $\alpha$-cluster case, followed by a modest decrease in $ v_{3}\{2, |\Delta\eta|>1.0\}/ v_{2}\{2, |\Delta\eta|>1.0\}$ as a function of multiplicity class. Since these effects originate from the initial state, as shown in Fig.~\ref{fig:eccratio}, the stronger increase in the value of $ v_{3}\{2, |\Delta\eta|>1.0\}/ v_{2}\{2, |\Delta\eta|>1.0\}$ in the 0--10\% multiplicity class reflects the presence of a tetrahedral arrangement of four $\alpha$-clusters inside the $^{16}$O nuclei, giving rise to a large intrinsic triangular flow. Although one finds $v_{2}\{2, |\Delta\eta|>1.0\}$ for $\alpha$-cluster density profile is larger compared to that of Woods-Saxon nuclear density profile, this tetrahedral structure raises $v_{3}\{2, |\Delta\eta|>1.0\}$ even higher for the high-multiplicity events, which is also consistent with transport models such as AMPT, shown in Ref.~\cite{Behera:2023nwj}. Thus, the observation of a sudden rise of $v_{3}\{2, |\Delta\eta|>1.0\}$ in tetrahedral $\alpha$-clustered case can lead to an observation of a spike in $ v_{3}\{2, |\Delta\eta|>1.0\}/ v_{2}\{2, |\Delta\eta|>1.0\}$ for the highest charged particle multiplicity events and consequently, the observation of which in O--O collisions may signify the presence of a $\alpha$-clustered nuclear geometry of $^{16}$O nuclei.

Figure~\ref{fig:v2fluctuation} (a) shows the evolution of elliptic flow using two- and four-particle Q-cumulant method as a function of multiplicity class for Woods-Saxon and $\alpha$-cluster nuclear density profiles in O--O collisions at $\sqrt{s_{\rm NN}}=7$ TeV using IPGlasma + MUSIC + iSS + UrQMD model. One finds a small multiplicity dependence for $v_{2}\{2, |\Delta\eta|>1.0\}$ for both the nuclear density profiles within 0--30\% multiplicity class. In contrast, using the four-particle cumulants method, we observe a significant charged particle multiplicity dependence of elliptic flow. The value of $v_{2}\{4\}$ for the $\alpha$-cluster case remains almost unchanged as a function of multiplicity class, the same trend as of $v_{2}\{2\}$. However, for the Woods-Saxon case, $v_{2}\{4\}$ maintains a consistent decreasing feature with the decrease in the particle multiplicity.
Figure~\ref{fig:v2fluctuation} (b) shows the mean value of elliptic flow ($\langle v_{2}\rangle$), evaluated using Eq.~\eqref{eq:meanv2}, for Woods-Saxon and $\alpha$-cluster nuclear density profiles in O--O collisions at $\sqrt{s_{\rm NN}}=7$ TeV using IPGlasma + MUSIC + iSS + UrQMD model. $\langle v_{2}\rangle$ is found to be larger for O--O collisions having an $\alpha$-cluster nuclear density profile as compared to a Woods-Saxon distribution. The multiplicity class dependence of $\langle v_{2}\rangle$ for both the nuclear density profiles is similar to $v_{2}\{4\}$, as $v_{2}\{2, |\Delta\eta|>1.0\}$ is nearly independent of multiplicity class in the 0--30\% region.
Figure~\ref{fig:v2fluctuation} (c) shows a multiplicity class dependence of elliptic flow fluctuation ($\sigma_{v_{2}}$) for Woods-Saxon and $\alpha$-cluster nuclear density profiles in O--O collisions at $\sqrt{s_{\rm NN}}=7$ TeV using IPGlasma + MUSIC + iSS + UrQMD model. One finds that $\sigma_{v_{2}}$ for Woods-Saxon and $\alpha$-cluster nuclear density profiles have complementing multiplicity dependence, i.e., when going from the 0--10\% to 20--30\% multiplicity classes, $\sigma_{v_{2}}$ for Woods-Saxon increases while for an $\alpha$-cluster nuclei, it shows a decreasing trend. This makes one of the interesting distinctions between the two nuclear density profiles with the study of flow fluctuations. In Pb--Pb collisions, due to the absence of the elliptic geometry in the initial state, a larger contribution to the elliptic flow in the most central collisions comes from elliptic flow fluctuations; however, towards the mid-central collisions, the contribution of elliptic flow-fluctuations to elliptic flow decreases, as the elliptic geometry of the nuclear participants leads the elliptic flow contribution~\cite{ALICE:2014dwt, ATLAS:2017rtr, ATLAS:2017hap}. Interestingly, the $\alpha$-cluster case of O--O collisions shows a similarity in the evolution of elliptic flow fluctuations with the Pb--Pb collisions, which is clearly different in O--O collisions with a Woods-Saxon nuclear density profile. The source of this opposite behavior of  $\sigma_{v_{2}}$ for both the nuclear density profiles can thus intrinsically be attributed to the difference in the nuclear geometry of the $^{16}$O nucleus. Interestingly, the multiplicity class dependence of $\langle v_{2}\rangle$ and $\sigma_{v_{2}}$ are complementary to each other for the Woods-Saxon case in O--O collisions. However, this correlation between $\langle v_{2}\rangle$ and $\sigma_{v_{2}}$ is not completely applicable for O--O collisions with an $\alpha$-cluster nuclear density profile. 
Figure~\ref{fig:v2fluctuation} (d) shows the evolution of relative elliptic flow fluctuation ($F(v_{2})$) as a function of multiplicity class in O--O collisions at $\sqrt{s_{\rm NN}}=7$ TeV. Interestingly, we find a similar multiplicity dependence of $F(v_{2})$ as found for $\sigma_{v_{2}}$. The multiplicity class dependence of $F(v_{2})$ in O--O collisions with Woods-Saxon nuclear distribution resembles a p--Pb collision~\cite{ALICE:2014dwt}. In contrast, for $\alpha$-cluster nuclear density profiles, $F(v_{2})$ is higher for 0--10\% multiplicity class and decreases with a decrease in particle multiplicity. This is an important observation where we find that a change in the initial nucleon distribution (Woods-Saxon versus $\alpha$-cluster) can cause a difference in the multiplicity class or collision centrality dependence of relative elliptic flow fluctuations for a hydrodynamically evolving system.


\section{Summary}
\label{sec4}

In this paper, we study the anisotropic flow coefficients in O--O collisions at $\sqrt{s_{\rm NN}}=7$ TeV using a hybrid of IPGlasma + MUSIC + iSS + UrQMD model and try to probe the final state effects of the presence of an exotic $\alpha$-cluster nuclear geometry in contrast to a uniform Woods-Saxon nuclear distribution inside a $^{16}$O nucleus. Here, we study the multiplicity class dependence of anisotropic flow coefficients. In addition, for the first time, we report studies of elliptic flow fluctuations with a motivation to distinguish O--O collisions having $\alpha$-cluster geometry from a regular Woods-Saxon distribution. The observation of large triangular flow in the highest multiplicity O--O collisions having an $\alpha$-clustered nuclei compared to a Woods-Saxon nuclear distribution is consistent with transport model predictions. In addition, we observe a significant multiplicity class dependence of relative elliptic flow fluctuations in both the nuclear distributions. With this study, it is clear that the presence of an $\alpha$-cluster nuclear geometry has a significant effect on both elliptic and triangular flow. However, we note a distinctive feature appears when observables related to fluctuations are studied, i.e., triangular flow and elliptic flow fluctuations. While probing a nuclear density profile is a subject matter of low-energy nuclear physics, it will be very interesting to find an appropriate observable that can be measured in TeV nuclear collisions to do the job. This makes the present study more interesting.


\section*{Acknowledgements}
S.~P. acknowledges the doctoral fellowship from UGC, Government of India. R.~S. sincerely acknowledges the DAE-DST, Government of India funding under the mega-science project – “Indian participation in the ALICE experiment at CERN” bearing Project No. SR/MF/PS-02/2021-IITI (E-37123). G.~G.~B. gratefully acknowledges the Hungarian National Research, Development and Innovation Office (NKFIH) under Contracts No. OTKA K135515, No. NKFIH  NEMZ\_KI-2022-00031, ``2024-1.2.5-TET-2024-00022" and Wigner Scientific Computing Laboratory (WSCLAB, the former Wigner GPU Laboratory). The authors gratefully acknowledge the MoU between IIT Indore and Wigner Research Centre for Physics (WRCP), Hungary, for the techno-scientific international cooperation. 

\appendix
\section*{Appendix}

\subsection{Multiplicity class versus impact parameter}
\label{app:impB}

\begin{figure}
\begin{center}
\includegraphics[scale=0.4]{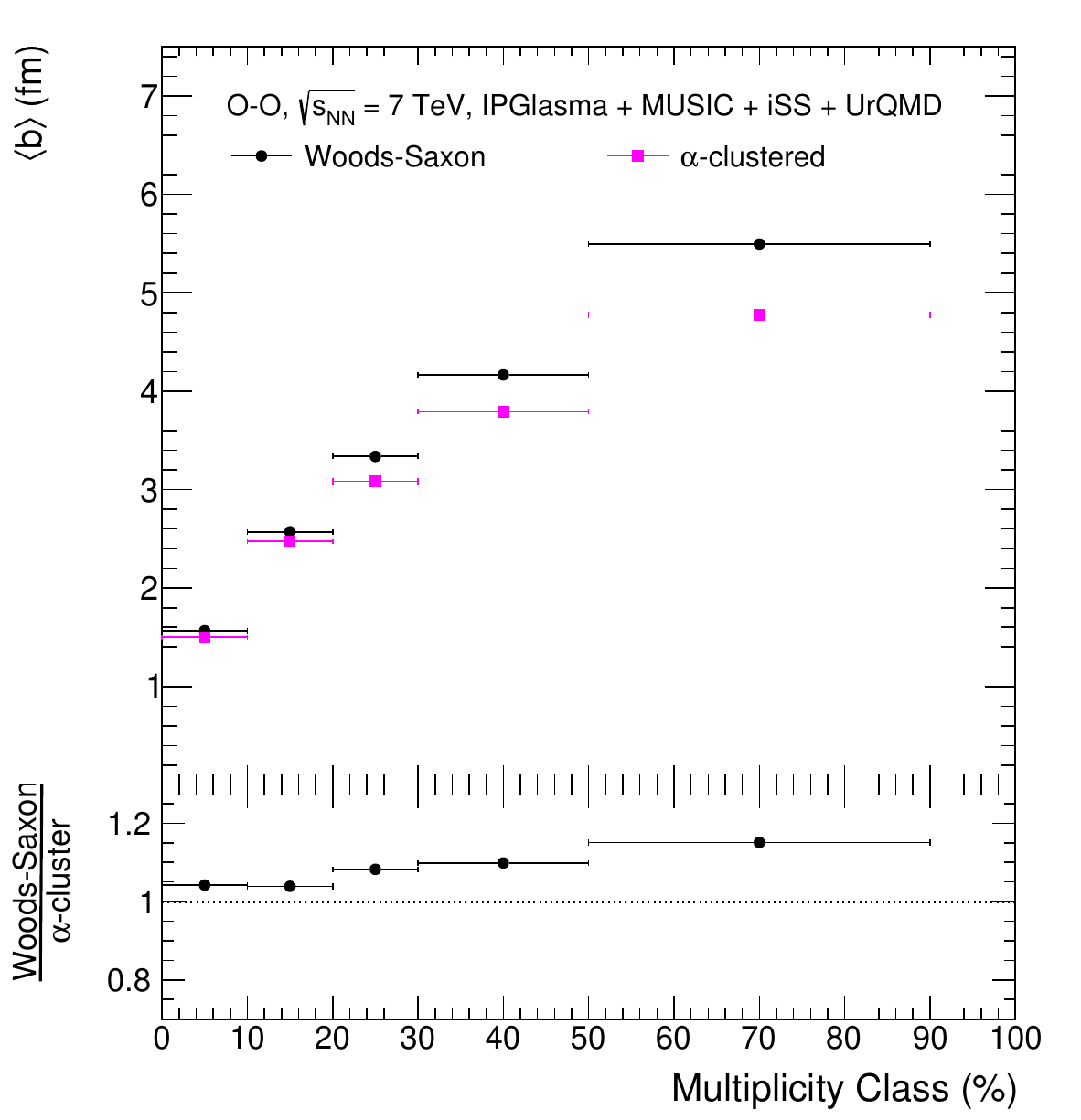}
\caption{Average value of impact parameter as a function of multiplicity class in O--O collisions at $\sqrt{s_{\rm NN}}=7$ TeV using IPGlasma + MUSIC + iSS + UrQMD for a Woods-Saxon and $\alpha$-cluster nuclear density profile.}
\label{fig:ImpBV0M}
\end{center}
\end{figure}

Figure~\ref{fig:ImpBV0M} shows the average value of the impact parameter ($\langle b\rangle$) as a function of multiplicity class in O--O collisions at $\sqrt{s_{\rm NN}}=7$ TeV using IPGlasma + MUSIC + iSS + UrQMD for Woods-Saxon and $\alpha$-cluster nuclear density profiles. As expected, one finds a significant dependence of $\langle b\rangle$ on event selection with the final state-charged particle multiplicity in the V0 acceptance region of the ALICE detector, i.e., the collisions having a lower impact parameter tend to produce a higher charged particle yield at the final state. In addition, one finds a significant dependence of $\langle b\rangle$ on the nuclear density profile of the colliding $^{16}$O nucleus to produce a similar charge particle yield in the final state. For the $\alpha$-cluster case, one finds that $\langle b\rangle$ is significantly smaller compared to a Woods-Saxon density profile because of the compact nature of $\alpha$-cluster nuclear structure, the effects of which are enhanced in the low multiplicity events. In addition, the compact nature of $\alpha$-cluster nuclear density profile is also reflected in the higher yield in 0--10\% multiplicity class, as shown in Fig.~\ref{fig:V0Mdist}.

\subsection{Eccentricity and triangularity}
\label{app:eccentricity}

\begin{figure}[!ht]
\begin{center}
\includegraphics[scale=0.4]{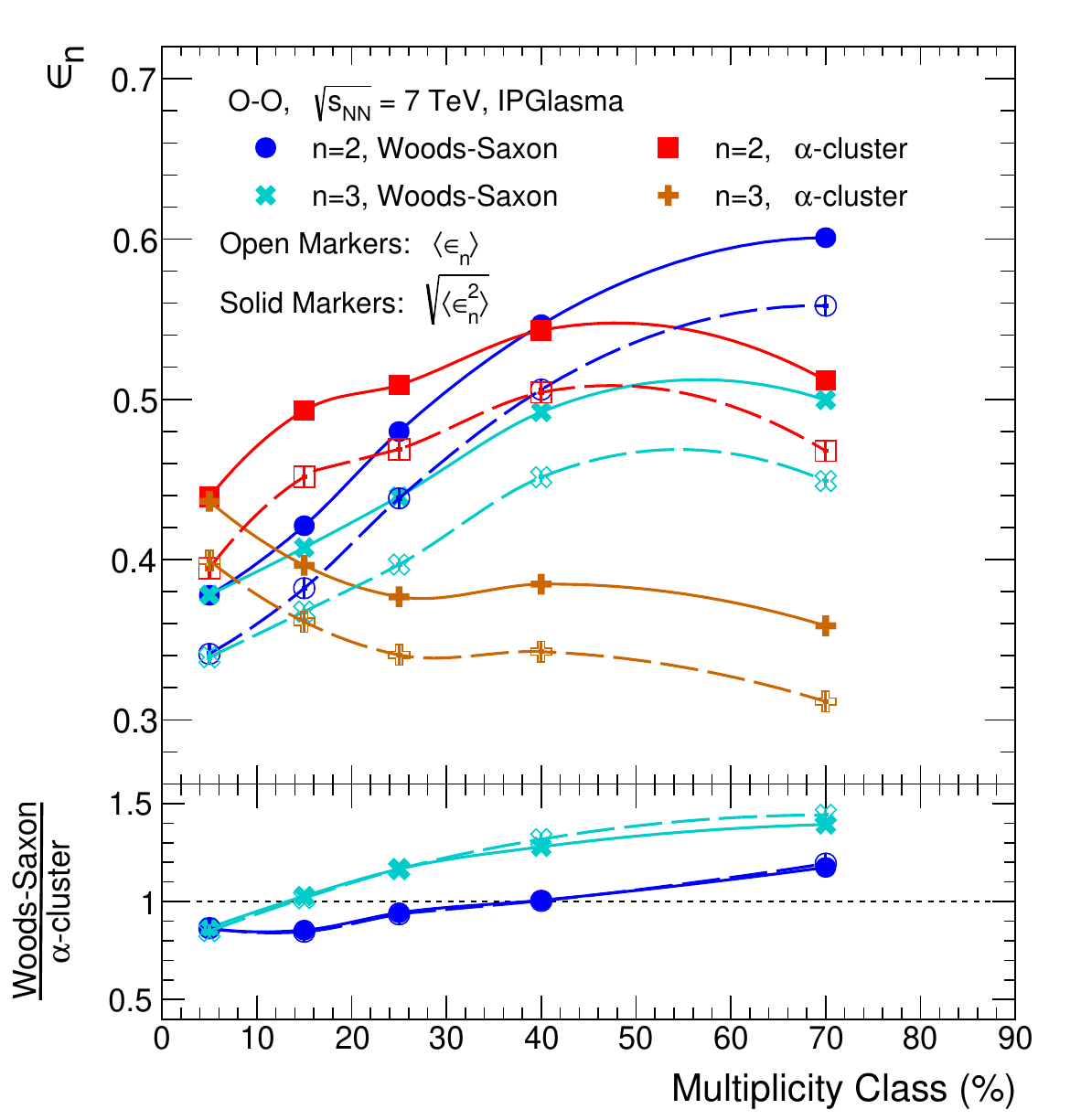}
\caption{Upper panel shows the average and RMS values of eccentricity and triangularity as a function of multiplicity class (\%) in O--O collisions at $\sqrt{s_{\rm NN}}=7$ TeV using IPGlasma for Woods-Saxon and $\alpha$-cluster nuclear density profiles. The lower panel shows the ratios of eccentricity and triangularity from Woods-Saxon to the $\alpha$-cluster nuclear density profile. In both the upper and lower panels, the vertical axes are zero-suppressed.}
\label{fig:ecc}
\end{center}
\end{figure}

The upper panel of Fig.~\ref{fig:ecc} shows the average eccentricity ($\langle\epsilon_2\rangle$), RMS eccentricity ($\sqrt{\langle\epsilon_{2}^{2}\rangle}$), average triangularity ($\langle\epsilon_3\rangle$), and RMS triangularity ($\sqrt{\langle\epsilon_{3}^{2}\rangle}$) as a function of multiplicity class in O--O collisions at $\sqrt{s_{\rm NN}}=7$ TeV using IPGlasma. Here, one finds that the average eccentricity for both the nuclear density profiles increases when one moves towards the lower multiplicity events, except for the $\alpha$-cluster case, which decreases slightly toward the 50--90\% multiplicity class. The rate of this increment with a decrease in charge particle multiplicity is lower for the $\alpha$-cluster case as compared to the Woods-Saxon nuclear density profile. One finds that the values of $\langle\epsilon_2\rangle$ for the $\alpha$-cluster case are higher than the corresponding values of Woods-Saxon nuclear density profiles for 0--40\% multiplicity classes; thereafter, Woods-Saxon has a higher value of $\langle\epsilon_2\rangle$. Further, one observes the average triangularity increases as one moves from the highest multiplicity region to the lower multiplicity region for the Woods-Saxon case. In contrast, $\langle\epsilon_3\rangle$ shows an interesting trend with the decrease in the charged particle multiplicity for the $\alpha$-cluster case. Both $\sqrt{\langle\epsilon_{2}^{2}\rangle}$ and $\sqrt{\langle\epsilon_{3}^{2}\rangle}$ retain a similar multiplicity class and nuclear density profile dependence observed for $\langle\epsilon_2\rangle$ and $\langle\epsilon_3\rangle$. It is important to note that average and RMS values of triangularity decrease by about 25\% and 20\%, respectively, when going from highest multiplicity to lowest multiplicity events of $\alpha$-cluster density profile.

The lower panel of Fig.~\ref{fig:ecc} shows the ratios of $\langle\epsilon_2\rangle$, $\langle\epsilon_3\rangle$, $\sqrt{\langle\epsilon_{2}^{2}\rangle}$, and $\sqrt{\langle\epsilon_{3}^{2}\rangle}$ in O--O collisions with Woods-Saxon nuclear density profile to $\alpha$-cluster case. Here, for the highest multiplicity events, i.e., 0--10\% multiplicity class, the Woods-Saxon nuclear density profile has smaller average and RMS values of $\epsilon_2$ and $\epsilon_3$ as compared to the corresponding values of $\alpha$-cluster case. Slowly, as the final state multiplicity decreases, the values of $\langle\epsilon_2\rangle$, $\langle\epsilon_3\rangle$, $\sqrt{\langle\epsilon_{2}^{2}\rangle}$, and $\sqrt{\langle\epsilon_{3}^{2}\rangle}$ gradually increase and become higher for the Woods-Saxon nuclear profile as compared to the corresponding values of $\alpha$-cluster case. Here, one finds that the ratio of average and RMS values of eccentricities and triangularities of the Woods-Saxon nuclear density profile to $\alpha$-cluster case are close to each other.

\begin{figure}
\begin{center}
\includegraphics[scale=0.4]{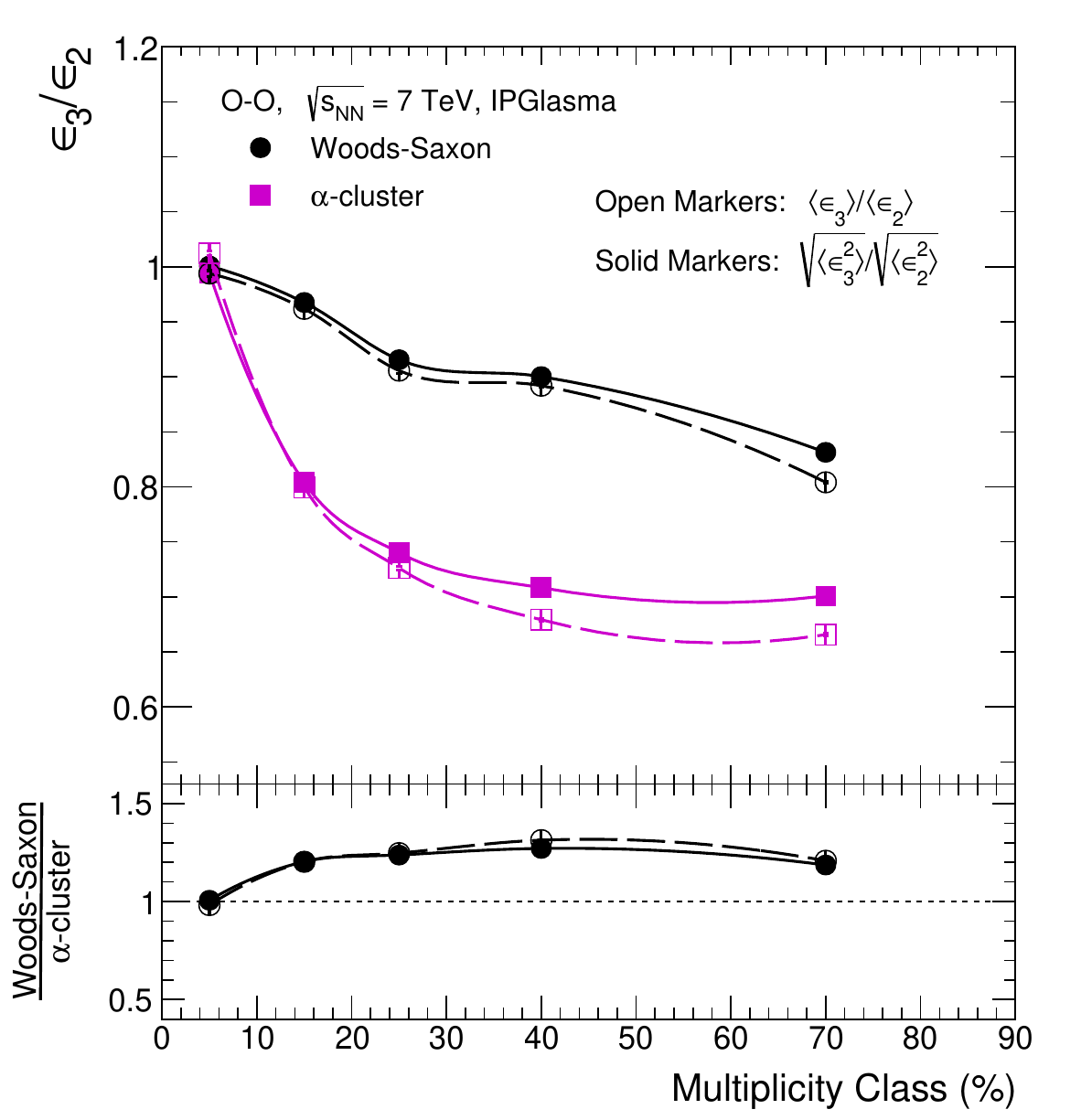}
\caption{Upper panel shows the ratios $\langle\epsilon_{3}\rangle/\langle\epsilon_{2}\rangle$ and $\sqrt{\langle\epsilon_{3}^{2}\rangle}/\sqrt{\langle\epsilon_{2}^{2}\rangle}$ as a function of multiplicity class (\%) in O--O collisions at $\sqrt{s_{\rm NN}}=7$ TeV using IPGlasma model for Woods-Saxon and $\alpha$-cluster nuclear density profiles. The lower panel shows the ratio from Woods-Saxon to the $\alpha$-cluster nuclear density profile. In both the upper and lower panels, the vertical axes are zero-suppressed.}
\label{fig:eccratio}
\end{center}
\end{figure}

The upper panel of Fig.~\ref{fig:eccratio} shows the ratios $\langle\epsilon_{3}\rangle/\langle\epsilon_{2}\rangle$ and $\sqrt{\langle\epsilon_{3}^{2}\rangle}/\sqrt{\langle\epsilon_{2}^{2}\rangle}$ as a function of multiplicity class (\%) in O--O collisions at $\sqrt{s_{\rm NN}}=7$ TeV using IPGlasma model for Woods-Saxon and $\alpha$-cluster nuclear density profiles. Here, one finds that, from 0--10\% to 10--20\% multiplicity class, the values of both $\langle\epsilon_{3}\rangle/\langle\epsilon_{2}\rangle$ and $\sqrt{\langle\epsilon_{3}^{2}\rangle}/\sqrt{\langle\epsilon_{2}^{2}\rangle}$ decrease rapidly for the $\alpha$-cluster case in contrast to Woods-Saxon nuclear density profile. Further, the values of $\langle\epsilon_{3}\rangle/\langle\epsilon_{2}\rangle$ and $\sqrt{\langle\epsilon_{3}^{2}\rangle}/\sqrt{\langle\epsilon_{2}^{2}\rangle}$ for the Woods-Saxon nuclear profile are higher as compared to the $\alpha$-cluster case except for the 0--10\% multiplicity class, as can be seen in the lower panel of Fig.~\ref{fig:eccratio}.

\subsection{Four particle cumulants}
\label{app:c4n}

\begin{figure}
\begin{center}
\includegraphics[scale=0.4]{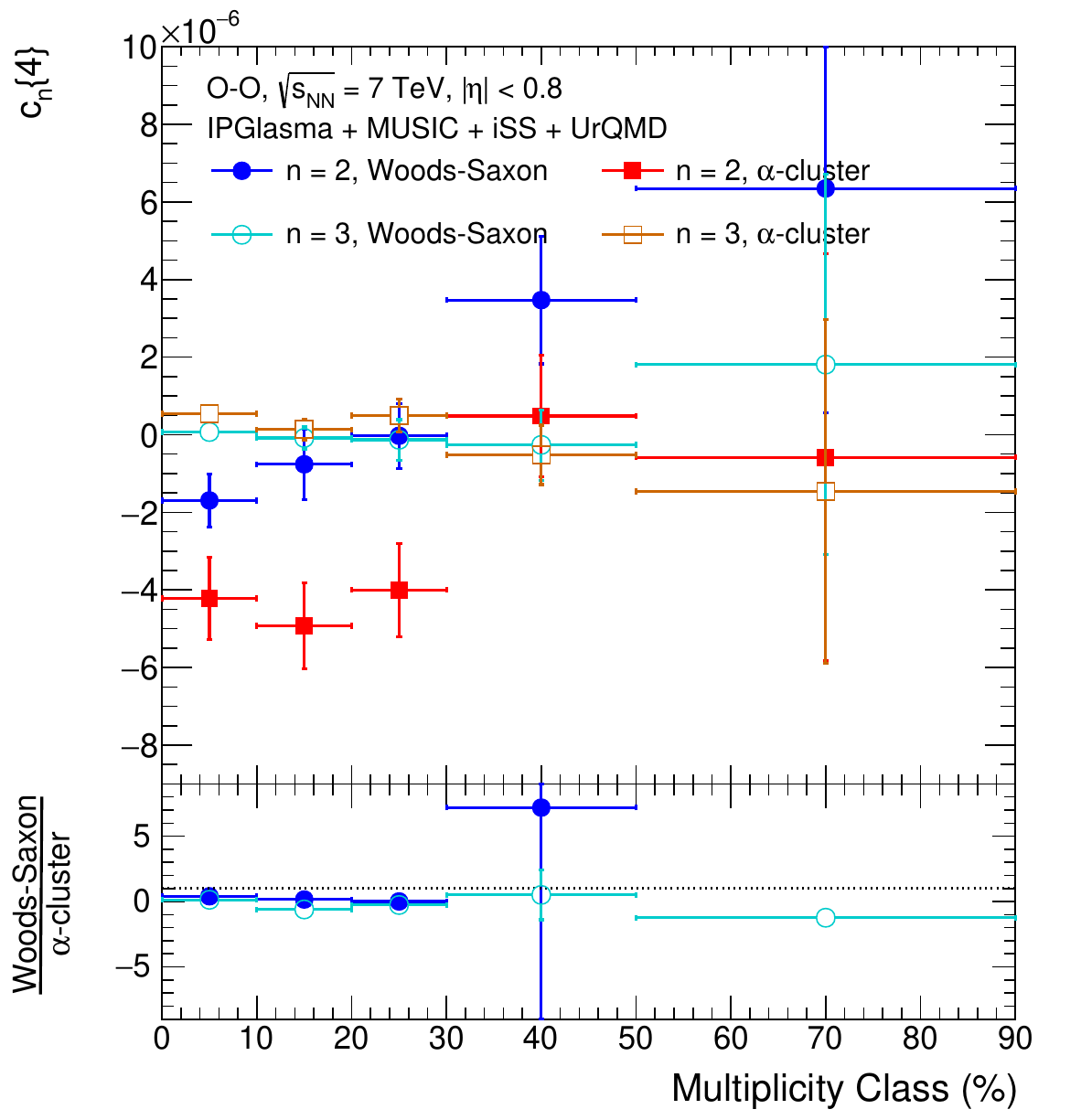}
\caption{Upper panel shows the second and third-order four-particle Q-cumulants ($c_{n}\{4\}$) as a function of multiplicity class (\%) in O--O collisions at $\sqrt{s_{\rm NN}}=7$ TeV using IPGlasma + MUSIC + iSS + UrQMD model for Woods-Saxon and $\alpha$-cluster nuclear density profiles. Lower panel shows the ratio of $c_{n}\{4\}$ from Woods-Saxon to a $\alpha$-cluster nuclear density profile.}
\label{fig:cn4}
\end{center}
\end{figure}

\begin{figure*}
\begin{center}
\includegraphics[scale=0.28]{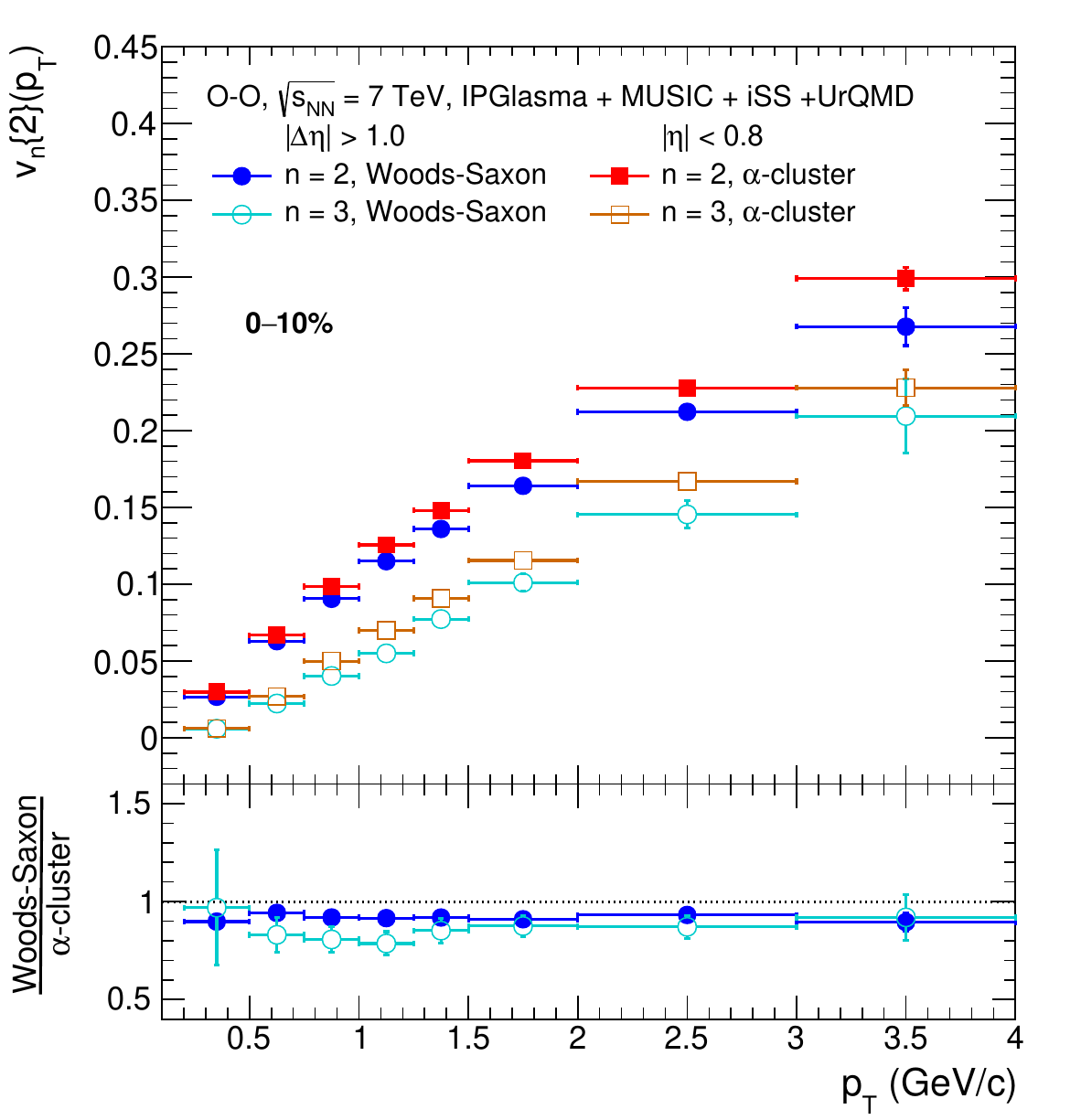}
\includegraphics[scale=0.28]{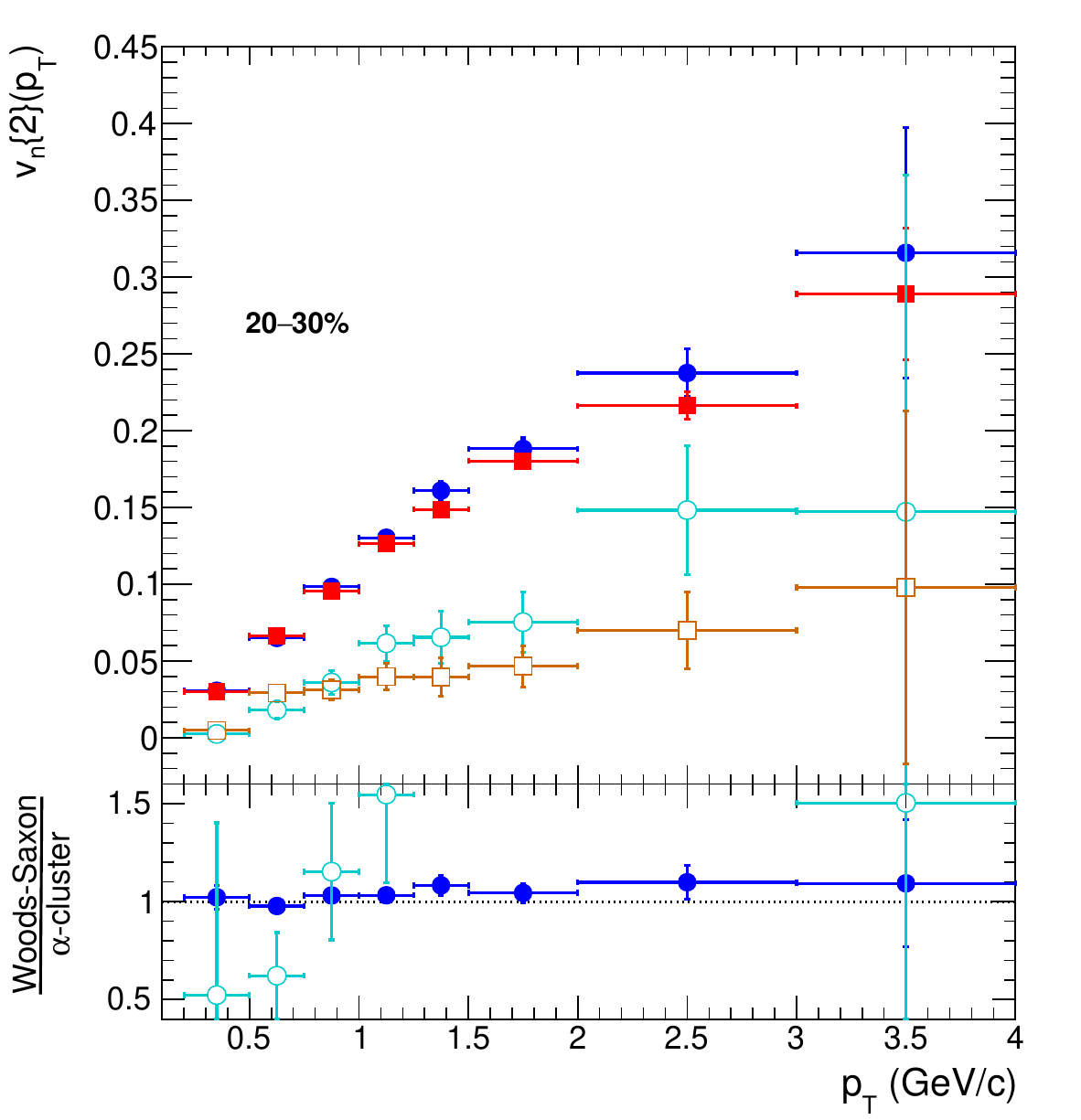}
 \includegraphics[scale=0.28]{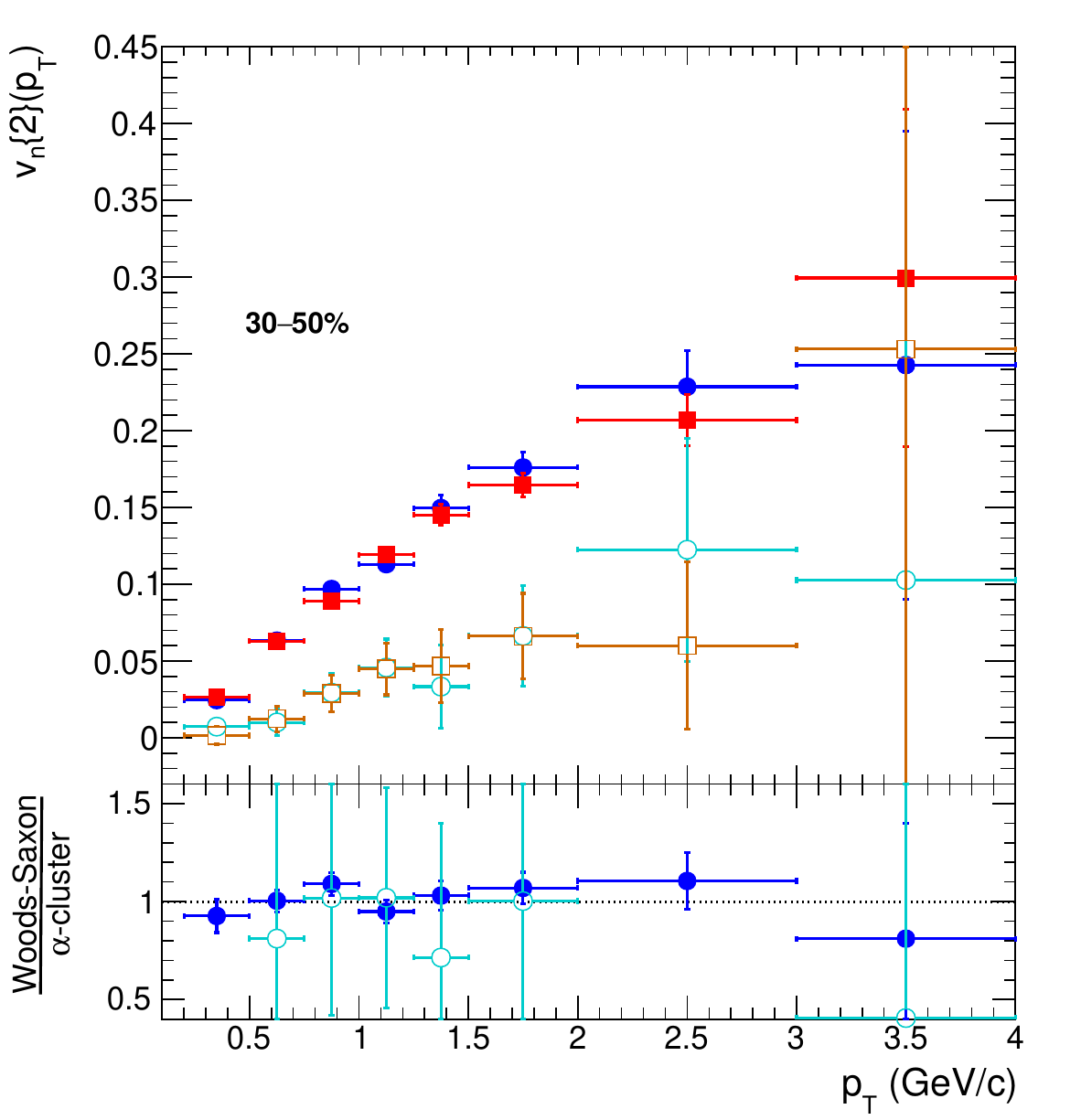}
\caption{Upper panel shows the elliptic flow and triangular flow using two-particle Q-cumulant method as a function of transverse momentum for different multiplicity classes in O--O collisions at $\sqrt{s_{\rm NN}}=7$ TeV using IPGlasma + MUSIC + iSS + UrQMD model for Woods-Saxon and $\alpha$-cluster nuclear density profiles. The lower panel shows the ratio of $v_{n}\{2\}(p_{\rm T})$ for the Woods-Saxon density profile to $\alpha$-cluster density profile.}
\label{fig:vnpT}
\end{center}
\end{figure*}

The upper panel of Fig.~\ref{fig:cn4} shows the multiplicity class dependence of four-particle cumulants ($c_{n}\{4\}$) in O--O collisions at $\sqrt{s_{\rm NN}}=7$ TeV using IPGlasma + MUSIC + iSS + UrQMD model for Woods-Saxon and $\alpha$-cluster nuclear density profiles. The lower panel depicts the ratio of $c_{n}\{4\}$ from Woods-Saxon to a $\alpha$-cluster nuclear density profile. One can note that, $v_{n}\{4\}$ is related with $c_{n}\{4\}$ using Eq.~\eqref{eq:vncnrelation}, thus a negative value of $c_{n}\{4\}$ results in a physically acceptable value of $v_{n}\{4\}$. However, a positive value of $c_{n}\{4\}$ leads to an imaginary value of $v_{n}\{4\}$, which sometimes occur when the system is dominated with non-flow contributions~\cite{ALICE:2014dwt}. In the upper panel of Fig.~\ref{fig:cn4}, one finds a significant difference of $c_{n}\{4\}$ between two nuclear density profiles. For an $\alpha$-cluster case, $c_{2}\{4\}$ starts with a small negative value in the 0--10\% multiplicity class and starts to decrease towards 10--20\% multiplicity class, attains a minimum and goes on increasing until it attains a positive value beyond 30--40\% multiplicity class. In contrast, $c_{2}\{4\}$ for a Woods-Saxon nuclear density profile starts at a minimum value for the most central case, i.e., 0--10\% multiplicity class, and starts to rise, becomes positive beyond 30--40\% multiplicity class. Interestingly, for both the nuclear density profiles, $c_{2}\{4\}$ attains a positive value beyond a similar 30--40\% multiplicity class, which naively implies the dominance of non-flow contribution for events with low final state multiplicity. The positive values of $c_{2}\{4\}$ on and beyond 30--40\% multiplicity class restrict us from the calculation of four-particle elliptic flow and its fluctuations below 30--40\% multiplicity class. On the other hand, $c_{3}\{4\}$ for both the nuclear density profiles fluctuate around 0, with a much smaller value as compared to corresponding $c_{2}\{4\}$. The positive values of $c_{3}\{4\}$ correspond to large non-flow azimuthal correlations.

\subsection{$p_{\rm T}$-dependence of anisotropic flow}
\label{app:pTvn}

Figure~\ref{fig:vnpT} shows the transverse momentum dependence of elliptic and triangular flow calculated using the two-particle Q-cumulants method in O--O collisions at $\sqrt{s_{\rm NN}}=7$ TeV using IPGlasma + MUSIC + iSS + UrQMD model for Woods-Saxon and $\alpha$-cluster nuclear density profiles in the upper panel. The bottom panel shows the ratio of $v_{n}\{2\}(p_{\rm T})$ from the Woods-Saxon density profile to $\alpha$-cluster density profile. One finds both $ v_{2}\{2, |\Delta\eta|>1.0\}(p_{\rm T})$ and $ v_{3}\{2, |\Delta\eta|>1.0\}(p_{\rm T})$ rise linearly with an increase in $p_{\rm T}$, which is an exclusive feature of hydro-based models. In addition, one observes that for the 0--10\% class, $ v_{2}\{2, |\Delta\eta|>1.0\}(p_{\rm T})$ and $ v_{3}\{2, |\Delta\eta|>1.0\}(p_{\rm T})$ are closer to one another, where the separation further increases as one moves towards 20--30\% and 30--50\% multiplicity classes. This is because the contributions for $ v_{2}\{2, |\Delta\eta|>1.0\}(p_{\rm T})$ and $ v_{3}\{2, |\Delta\eta|>1.0\}(p_{\rm T})$ are the consequences of the interplay between the geometry of the overlap region and the eccentricity fluctuations during the collisions, etc., where the eccentricity fluctuations dominate the contributions to the anisotropic flow in the central collisions. Another interesting observation is that in 0--10\% multiplicity class, $v_{n}\{2\}(p_{\rm T})$ for the $\alpha$-cluster case is larger compared to corresponding values from a Woods-Saxon distribution. This observation is consistent with the Fig~\ref{fig:vn2etagap}. However, for 20--30\% and 30--50\% multiplicity classes, $v_{n}\{2\}(p_{\rm T})$ from Woods-Saxon nuclear distribution is larger compared to an $\alpha$-cluster nuclear density profile. For the 0--10\% multiplicity class, the lower ratio plot of Woods-Saxon to an $\alpha$-cluster nuclear density profile for both elliptic and triangular flow shows a flat ratio, almost independent of $p_{\rm T}$. However, it no longer holds towards the lower multiplicity classes, indicating a $p_{\rm T}$ dependence of medium effects for different nuclear profiles to the anisotropic flow coefficients.


\section*{Software}
In this work, we used publicly available software for data generation. The software for IPGlasma, MUSIC, iSS and UrQMD can be accessed using the web links provided in Refs.~\cite{IPGlasmaRef, MUSICRef, iSSRef, UrQMDRef}, respectively.


\newpage

\begin{thebibliography}{100}

\bibitem{Voloshin:1994mz}
S.~Voloshin and Y.~Zhang,
Z. Phys. C \textbf{70}, 665 (1996).

\bibitem{Schenke:2011bn}
B.~Schenke, S.~Jeon and C.~Gale,
Phys. Rev. C \textbf{85}, 024901 (2012).

\bibitem{Chaudhuri:2011pa}
A.~K.~Chaudhuri,
Phys. Lett. B \textbf{713}, 91 (2012)


\bibitem{Giacalone:2021udy}
G.~Giacalone, J.~Jia and C.~Zhang,
Phys. Rev. Lett. \textbf{127}, 242301 (2021).

\bibitem{Haque:2019vgi}
M.~R.~Haque, M.~Nasim and B.~Mohanty,
J. Phys. G \textbf{46}, 085104 (2019).

\bibitem{Behera:2023nwj}
D.~Behera, S.~Prasad, N.~Mallick and R.~Sahoo,
Phys. Rev. D \textbf{108}, 054022 (2023).

\bibitem{ALICE:2021ibz}
S.~Acharya \textit{et al.} (ALICE Collaboration),
JHEP \textbf{10}, 152 (2021).

\bibitem{ALICE:2018lao}
S.~Acharya \textit{et al.} (ALICE Collaboration),
Phys. Lett. B \textbf{784}, 82 (2018).

\bibitem{CMS:2019cyz}
A.~M.~Sirunyan \textit{et al.} (CMS Collaboration),
Phys. Rev. C \textbf{100}, 044902 (2019).

\bibitem{ATLAS:2019dct}
G.~Aad \textit{et al.} (ATLAS Collaboration),
Phys. Rev. C \textbf{101}, 024906 (2020).

\bibitem{STAR:2015mki}
L.~Adamczyk \textit{et al.} (STAR Collaboration),
Phys. Rev. Lett. \textbf{115}, no.22, 222301 (2015).

\bibitem{PHENIX:2018lia}
C.~Aidala \textit{et al.} (PHENIX Collaboration),
Nature Phys. \textbf{15}, 214 (2019).

\bibitem{Gardim:2011xv}
F.~G.~Gardim, F.~Grassi, M.~Luzum and J.~Y.~Ollitrault,
Phys. Rev. C \textbf{85}, 024908 (2012).

\bibitem{ALICE:2022zks}
S.~Acharya \textit{et al.} (ALICE Collaboration),
JHEP \textbf{05}, 243 (2023).

\bibitem{CMS:2015xmx}
V.~Khachatryan \textit{et al.} (CMS Collaboration),
Phys. Rev. C \textbf{92}, 034911 (2015).

\bibitem{ALICE:2017lyf}
S.~Acharya \textit{et al.} (ALICE Collaboration),
JHEP \textbf{09}, 032 (2017).

\bibitem{ATLAS:2017rij}
M.~Aaboud \textit{et al.} (ATLAS Collaboration),
Eur. Phys. J. C \textbf{78}, 142 (2018).

\bibitem{PHOBOS:2007vdf}
B.~Alver \textit{et al.} (PHOBOS Collaboration),
Phys. Rev. C \textbf{77}, 014906 (2008).

\bibitem{PHOBOS:2010ekr}
B.~Alver \textit{et al.} (PHOBOS Collaboration),
Phys. Rev. C \textbf{81}, 034915 (2010).

\bibitem{PHENIX:2018lfu}
A.~Adare \textit{et al.} (PHENIX Collaboration),
Phys. Rev. C \textbf{99}, 024903 (2019).

\bibitem{ALICE:2018yph}
S.~Acharya \textit{et al.} (ALICE Collaboration),
JHEP \textbf{09}, 006 (2018).


\bibitem{ALICE:2012vgf}
B.~Abelev \textit{et al.} (ALICE Collaboration),
Phys. Lett. B \textbf{719}, 18 (2013).

\bibitem{ALICE:2014wao}
B.~B.~Abelev \textit{et al.} (ALICE Collaboration),
JHEP \textbf{06}, 190 (2015).

\bibitem{ALICE:2016cti}
J.~Adam \textit{et al.} (ALICE Collaboration),
JHEP \textbf{09}, 164 (2016).

\bibitem{ATLAS:2013xzf}
G.~Aad \textit{et al.} (ATLAS Collaboration),
JHEP \textbf{11}, 183 (2013).


\bibitem{CMS:2017glf}
A.~M.~Sirunyan \textit{et al.} (CMS Collaboration),
Phys. Lett. B \textbf{789}, 643 (2019).


\bibitem{ALICE:2016fzo}
J.~Adam \textit{et al.} (ALICE Collaboration),
Nature Phys. \textbf{13}, 535 (2017).

\bibitem{ALICE:2013snk}
B.~B.~Abelev \textit{et al.} (ALICE Collaboration),
Phys. Lett. B \textbf{726}, 164 (2013).


\bibitem{CMS:2015fgy}
V.~Khachatryan \textit{et al.} (CMS Collaboration),
Phys. Rev. Lett. \textbf{116}, 172302 (2016).

\bibitem{ALICE:2013wgn}
B.~B.~Abelev \textit{et al.} (ALICE Collaboration),
Phys. Lett. B \textbf{728}, 25 (2014).

\bibitem{ALICE:2016dei}
J.~Adam \textit{et al.} (ALICE Collaboration),
Phys. Lett. B \textbf{760}, 720 (2016).

\bibitem{CMS:2016fnw}
V.~Khachatryan \textit{et al.} (CMS Collaboration),
Phys. Lett. B \textbf{765}, 193 (2017).

\bibitem{ALICE:2024vzv}
S.~Acharya \textit{et al.} (ALICE Collaboration),
[arXiv:2411.09323 [nucl-ex]].

\bibitem{Brewer:2021kiv}
J.~Brewer, A.~Mazeliauskas and W.~van der Schee,
[arXiv:2103.01939 [hep-ph]].

\bibitem{Katz:2019qwv}
R.~Katz, C.~A.~G.~Prado, J.~Noronha-Hostler and A.~A.~P.~Suaide,
Phys. Rev. C \textbf{102}, 041901 (2020).


\bibitem{gamow}
G.~Gamow, Constitution of Atomic Nuclei and Radioactivity,
International series of monographs on physics PCMI collection, Clarendon Press (1931).

\bibitem{Wheeler:1937zza}
J.~A.~Wheeler,
Phys. Rev. \textbf{52}, 1083 (1937).


\bibitem{Bijker:2014tka}
R.~Bijker and F.~Iachello,
Phys. Rev. Lett. \textbf{112}, 152501 (2014).

\bibitem{Wang:2019dpl}
X.~B.~Wang, G.~X.~Dong, Z.~C.~Gao, Y.~S.~Chen and C.~W.~Shen,
Phys. Lett. B \textbf{790}, 498 (2019).

\bibitem{He:2014iqa}
W.~B.~He, Y.~G.~Ma, X.~G.~Cao, X.~Z.~Cai and G.~Q.~Zhang,
Phys. Rev. Lett. \textbf{113}, 032506 (2014).

\bibitem{He:2021uko}
J.~He, W.~B.~He, Y.~G.~Ma and S.~Zhang,
Phys. Rev. C \textbf{104}, 044902 (2021).

\bibitem{Otsuka:2022bcf}
T.~Otsuka, T.~Abe, T.~Yoshida, Y.~Tsunoda, N.~Shimizu, N.~Itagaki, Y.~Utsuno, J.~Vary, P.~Maris and H.~Ueno,
Nature Commun. \textbf{13}, 2234 (2022).


\bibitem{PHENIX:2021ubk}
U.~A.~Acharya \textit{et al.} (PHENIX Collaboration),
Phys. Rev. C \textbf{105}, 024901 (2022).

\bibitem{STAR:2022pfn}
M.~I.~Abdulhamid \textit{et al.} (STAR Collaboration),
Phys. Rev. Lett. \textbf{130}, 242301 (2023).

\bibitem{STAR:2023wmd}
 (STAR Collaboration),
[arXiv:2312.07464 [nucl-ex]].

\bibitem{Rybczynski:2019adt}
M.~Rybczy\'nski and W.~Broniowski,
Phys. Rev. C \textbf{100}, 064912 (2019).

\bibitem{Sievert:2019zjr}
M.~D.~Sievert and J.~Noronha-Hostler,
Phys. Rev. C \textbf{100}, 024904 (2019).

\bibitem{Huang:2019tgz}
S.~Huang, Z.~Chen, J.~Jia and W.~Li,
Phys. Rev. C \textbf{101}, 021901 (2020).

\bibitem{Behera:2021zhi}
D.~Behera, N.~Mallick, S.~Tripathy, S.~Prasad, A.~N.~Mishra and R.~Sahoo,
Eur. Phys. J. A \textbf{58}, 175 (2022).

\bibitem{Li:2020vrg}
Y.~A.~Li, S.~Zhang and Y.~G.~Ma,
Phys. Rev. C \textbf{102}, 054907 (2020).

\bibitem{Bozek:2014cva}
P.~Bozek, W.~Broniowski, E.~Ruiz Arriola and M.~Rybczynski,
Phys. Rev. C \textbf{90}, 064902 (2014).

\bibitem{Broniowski:2013dia}
W.~Broniowski and E.~Ruiz Arriola,
Phys. Rev. Lett. \textbf{112}, 112501 (2014).

\bibitem{Giacalone:2024luz}
G.~Giacalone, \textit{et al.}
[arXiv:2402.05995 [nucl-th]].

\bibitem{Lim:2018huo}
S.~H.~Lim, J.~Carlson, C.~Loizides, D.~Lonardoni, J.~E.~Lynn, J.~L.~Nagle, J.~D.~Orjuela Koop and J.~Ouellette,
Phys. Rev. C \textbf{99}, 044904 (2019).

\bibitem{Summerfield:2021oex}
N.~Summerfield, B.~N.~Lu, C.~Plumberg, D.~Lee, J.~Noronha-Hostler and A.~Timmins,
Phys. Rev. C \textbf{104}, L041901 (2021).

\bibitem{Schenke:2020mbo}
B.~Schenke, C.~Shen and P.~Tribedy,
Phys. Rev. C \textbf{102}, 044905 (2020).


\bibitem{Huss:2020whe}
A.~Huss, A.~Kurkela, A.~Mazeliauskas, R.~Paatelainen, W.~van der Schee and U.~A.~Wiedemann,
Phys. Rev. C \textbf{103}, 054903 (2021).

\bibitem{Zakharov:2021uza}
B.~G.~Zakharov,
JHEP \textbf{09}, 087 (2021).

\bibitem{Behera:2023oxe}
D.~Behera, S.~Deb, C.~R.~Singh and R.~Sahoo,
Phys. Rev. C \textbf{109}, 014902 (2024).

\bibitem{Ding:2023ibq}
C.~Ding, L.~G.~Pang, S.~Zhang and Y.~G.~Ma,
Chin. Phys. C \textbf{47}, 024105 (2023).

\bibitem{Wang:2021ghq}
Y.~Z.~Wang, S.~Zhang and Y.~G.~Ma,
Phys. Lett. B \textbf{831}, 137198 (2022).

\bibitem{Rybczynski:2017nrx}
M.~Rybczy\'nski, M.~Piotrowska and W.~Broniowski,
Phys. Rev. C \textbf{97}, 034912 (2018).

\bibitem{Svetlichnyi:2023nim}
A.~Svetlichnyi, S.~Savenkov, R.~Nepeivoda and I.~Pshenichnov,
MDPI Physics \textbf{5}, 381 (2023).


\bibitem{Giacalone:2024ixe}
G.~Giacalone, \textit{et al.}
[arXiv:2405.20210 [nucl-th]].


\bibitem{Zhang:2024vkh}
C.~Zhang, J.~Chen, G.~Giacalone, S.~Huang, J.~Jia and Y.~G.~Ma,
[arXiv:2404.08385 [nucl-th]].

\bibitem{R:2024eni}
A.~M.~K.~R, S.~Prasad, N.~Mallick and R.~Sahoo,
[arXiv:2407.03823 [nucl-th]].

\bibitem{Lin:2004en}
Z.~W.~Lin, C.~M.~Ko, B.~A.~Li, B.~Zhang and S.~Pal,
Phys. Rev. C \textbf{72}, 064901 (2005).

\bibitem{McDonald:2016vlt}
S.~McDonald, C.~Shen, F.~Fillion-Gourdeau, S.~Jeon and C.~Gale,
Phys. Rev. C \textbf{95}, 064913 (2017).

\bibitem{Schenke:2012wb}
B.~Schenke, P.~Tribedy and R.~Venugopalan,
Phys. Rev. Lett. \textbf{108}, 252301 (2012).

\bibitem{Schenke:2012hg}
B.~Schenke, P.~Tribedy and R.~Venugopalan,
Phys. Rev. C \textbf{86}, 034908 (2012).

\bibitem{McLerran:1993ni}
L.~D.~McLerran and R.~Venugopalan,
Phys. Rev. D \textbf{49}, 2233 (1994).

\bibitem{McLerran:1993ka}
L.~D.~McLerran and R.~Venugopalan,
Phys. Rev. D \textbf{49}, 3352 (1994).

\bibitem{Bartels:2002cj}
J.~Bartels, K.~J.~Golec-Biernat and H.~Kowalski,
Phys. Rev. D \textbf{66}, 014001 (2002).

\bibitem{Kowalski:2003hm}
H.~Kowalski and D.~Teaney,
Phys. Rev. D \textbf{68}, 114005 (2003).

\bibitem{Loizides:2014vua}
C.~Loizides, J.~Nagle and P.~Steinberg,
SoftwareX \textbf{1-2}, 13 (2015).

\bibitem{McLerran:1994vd}
L.~D.~McLerran and R.~Venugopalan,
Phys. Rev. D \textbf{50}, 2225 (1994).

\bibitem{Krasnitz:1998ns}
A.~Krasnitz and R.~Venugopalan,
Nucl. Phys. B \textbf{557}, 237 (1999).

\bibitem{Schenke:2010nt}
B.~Schenke, S.~Jeon and C.~Gale,
Phys. Rev. C \textbf{82}, 014903 (2010).

\bibitem{Schenke:2010rr}
B.~Schenke, S.~Jeon and C.~Gale,
Phys. Rev. Lett. \textbf{106}, 042301 (2011).

\bibitem{Paquet:2015lta}
J.~F.~Paquet, C.~Shen, G.~S.~Denicol, M.~Luzum, B.~Schenke, S.~Jeon and C.~Gale,
Phys. Rev. C \textbf{93}, 044906 (2016).

\bibitem{Huovinen:2009yb}
P.~Huovinen and P.~Petreczky,
Nucl. Phys. A \textbf{837}, 26 (2010).

\bibitem{Kurganov:2000ovy}
A.~Kurganov and E.~Tadmor,
J. Comput. Phys. \textbf{160}, 241 (2000).

\bibitem{Jeon:2015dfa}
S.~Jeon and U.~Heinz,
Int. J. Mod. Phys. E \textbf{24}, 1530010 (2015).

\bibitem{Shen:2014vra}
C.~Shen, Z.~Qiu, H.~Song, J.~Bernhard, S.~Bass and U.~Heinz,
Comput. Phys. Commun. \textbf{199}, 61 (2016).

\bibitem{Denicol:2018wdp}
G.~S.~Denicol, C.~Gale, S.~Jeon, A.~Monnai, B.~Schenke and C.~Shen,
Phys. Rev. C \textbf{98}, 034916 (2018).

\bibitem{Cooper:1974mv}
F.~Cooper and G.~Frye,
Phys. Rev. D \textbf{10}, 186 (1974).

\bibitem{Dusling:2009df}
K.~Dusling, G.~D.~Moore and D.~Teaney,
Phys. Rev. C \textbf{81}, 034907 (2010).


\bibitem{Bass:1998ca}
S.~A.~Bass, M.~Belkacem, M.~Bleicher, M.~Brandstetter, L.~Bravina, C.~Ernst, L.~Gerland, M.~Hofmann, S.~Hofmann and J.~Konopka, \textit{et al.}
Prog. Part. Nucl. Phys. \textbf{41}, 255 (1998).

\bibitem{Bleicher:1999xi}
M.~Bleicher, E.~Zabrodin, C.~Spieles, S.~A.~Bass, C.~Ernst, S.~Soff, L.~Bravina, M.~Belkacem, H.~Weber and H.~Stoecker, \textit{et al.}
J. Phys. G \textbf{25}, 1859 (1999).


\bibitem{Mallick:2023vgi}
N.~Mallick, S.~Prasad, A.~N.~Mishra, R.~Sahoo and G.~G.~Barnaf\"oldi,
Phys. Rev. D \textbf{107}, 094001 (2023).

\bibitem{Mallick:2022alr}
N.~Mallick, S.~Prasad, A.~N.~Mishra, R.~Sahoo and G.~G.~Barnaf\"oldi,
Phys. Rev. D \textbf{105}, 114022 (2022).

\bibitem{Prasad:2022zbr}
S.~Prasad, N.~Mallick, S.~Tripathy and R.~Sahoo,
Phys. Rev. D \textbf{107}, 074011 (2023).



\bibitem{Bilandzic:2010jr}
A.~Bilandzic, R.~Snellings and S.~Voloshin,
Phys. Rev. C \textbf{83}, 044913 (2011).

\bibitem{Bilandzic:2013kga}
  A.~Bilandzic, C.~H.~Christensen, K.~Gulbrandsen, A.~Hansen and Y.~Zhou,
  Phys.\ Rev.\ C {\bf 89}, 064904 (2014).
  
  

\bibitem{Aamodt:2010pa}
  KAamodt {\it et al.}  (ALICE Collaboration),
  Phys.\ Rev.\ Lett.\  {\bf 105}, 252302 (2010).

\bibitem{ALICE:2011ab}
  K.~Aamodt {\it et al.}  (ALICE Collaboration),
  Phys.\ Rev.\ Lett.\  {\bf 107}, 032301 (2011);
  
\bibitem{Zhou:2014bba}
Y.~Zhou (ALICE Collaboration),
Nucl. Phys. A \textbf{931}, 949 (2014).
  

\bibitem{ALICE:2014dwt}
B.~B.~Abelev \textit{et al.} (ALICE Collaboration),
Phys. Rev. C \textbf{90}, 054901 (2014).

\bibitem{Zhou:2015iba}
Y.~Zhou, X.~Zhu, P.~Li and H.~Song,
Phys. Rev. C \textbf{91}, 064908 (2015).

\bibitem{STAR:2002hbo}
C.~Adler \textit{et al.} (STAR Collaboration),
Phys. Rev. C \textbf{66}, 034904 (2002).

\bibitem{Jia:2017hbm}
J.~Jia, M.~Zhou and A.~Trzupek,
Phys. Rev. C \textbf{96}, 034906 (2017).

\bibitem{ATLAS:2017rtr}
M.~Aaboud \textit{et al.} (ATLAS Collaboration),
Phys. Rev. C \textbf{97}, 024904 (2018).

\bibitem{ATLAS:2017hap}
M.~Aaboud \textit{et al.} (ATLAS Collaboration),
Eur. Phys. J. C \textbf{77}, 428 (2017).


\bibitem{Ollitrault:2009ie}
J.~Y.~Ollitrault, A.~M.~Poskanzer and S.~A.~Voloshin,
Phys. Rev. C \textbf{80}, 014904 (2009).

\bibitem{Bilandzic:2012wva}
A.~Bilandzic,
CERN-THESIS-2012-018.

\bibitem{IPGlasmaRef}
Available Online: \url{https://github.com/schenke/ipglasma/}.

\bibitem{MUSICRef}
Available Online: \url{https://github.com/MUSIC-fluid/MUSIC/}.

\bibitem{iSSRef}
Available Online: \url{https://github.com/chunshen1987/iSS/}.

\bibitem{UrQMDRef}
Available Online: \url{https://itp.uni-frankfurt.de/~bleicher/index.html?content=urqmd}.

\end{thebibliography}
\end{document}